\documentclass[prl,aps,twocolumn,floats,nofootinbib,superscriptaddress]{revtex4-1}
\usepackage{amsmath,amssymb,graphicx,psfrag}
\usepackage[colorlinks,linkcolor={blue},citecolor={blue},urlcolor={blue}]{hyperref}

\begin{document}
\renewcommand{\ni}{{\noindent}}
\newcommand{\dprime}{{\prime\prime}}
\newcommand{\be}{\begin{equation}}
\newcommand{\ee}{\end{equation}}
\newcommand{\bea}{\begin{eqnarray}} 
\newcommand{\eea}{\end{eqnarray}}
\newcommand{\la}{\langle}
\newcommand{\ra}{\rangle} 

\newcommand{\dg}{\dagger}
\newcommand\lbs{\left[}
\newcommand\rbs{\right]}
\newcommand\lbr{\left(}
\newcommand\rbr{\right)}
\newcommand\f{\frac}
\newcommand\e{\epsilon}
\newcommand\ua{\uparrow}
\newcommand\da{\downarrow}
\newcommand{\bcen}{\begin{center}}
\newcommand{\ecen}{\end{center}}
\newcommand{\btab}{\begin{tabular}}
\newcommand{\etab}{\end{tabular}}
\newcommand{\bdes}{\begin{description}}
\newcommand{\edes}{\end{description}}
\newcommand{\mc}{\multicolumn}
\newcommand{\ul}{\underline}
\newcommand{\non}{\nonumber}
\newcommand{\etal}{et.~al.\ }
\newcommand{\half}{\frac{1}{2}}
\newcommand{\bary}{\begin{array}}
\newcommand{\eary}{\end{array}}
\newcommand{\benum}{\begin{enumerate}}
\newcommand{\eenum}{\end{enumerate}}
\newcommand{\bitem}{\begin{itemize}}
\newcommand{\eitem}{\end{itemize}}
\newcommand{\cuup}[1]{c_{#1 \uparrow}}
\newcommand{\cdown}[1]{c_{#1 \downarrow}}
\newcommand{\cdup}[1]{c^\dagger_{#1 \uparrow}}
\newcommand{\cddown}[1]{c^\dagger_{#1 \downarrow}}
%
%
\newcommand{\beps}{\mbox{\boldmath $ \epsilon $}}
\newcommand{\bsig}{\mbox{\boldmath $ \sigma $}}
\newcommand{\bpi}{\mbox{\boldmath $ \pi $}}
\newcommand{\bkap}{\mbox{\boldmath $ \kappa $}}
\newcommand{\bgam}{\mbox{\boldmath $ \gamma $}}
\newcommand{\bphi}{\mbox{\boldmath $ \phi $}}
\newcommand{\balp}{\mbox{\boldmath $ \alpha $}}
\newcommand{\beot}{\mbox{\boldmath $ \eta $}}
\newcommand{\btau}{\mbox{\boldmath $ \tau $}}
\newcommand{\blam}{\mbox{\boldmath $ \lambda $}}
\newcommand{\bomg}{\mbox{\boldmath $ \omega $}}
\newcommand{\bOmg}{\mbox{\boldmath $ \Omega $}}
\newcommand{\bxhi}{\mbox{\boldmath $ \xi $}}
\newcommand{\bmu} {\mbox{\boldmath $ \mu $}}
\newcommand{\bnu} {\mbox{\boldmath $ \nu $}}
\newcommand{\bdelta}{{\boldsymbol{\delta}}}
\newcommand{\bTheta}{{\boldsymbol{\Theta}}}
\newcommand{\bpsi}{\mbox{\boldmath $ \psi $}}
\newcommand{\brho}{\mbox{\boldmath $ \rho $}}
\newcommand{\bGam}{\mbox{\boldmath $ \Gamma $}}
\newcommand{\bLam}{\mbox{\boldmath $ \Lambda $}}
\newcommand{\bPhi}{\mbox{\boldmath $ \Phi $}}
%
%
\newcommand{\ba} { \bm{a} }
\newcommand{\bb} { \mbox{\boldmath $b$}}
\newcommand{\bc} { {\mathbf c} }
\newcommand{\bd} { \mbox{\boldmath $d$}}
\newcommand{\bff}{ \mbox{\boldmath $f$}}
\newcommand{\bg} { \mbox{\boldmath $g$}}
\newcommand{\bh} { \mbox{\boldmath $h$}}
\newcommand{\bi} { \mbox{\boldmath $i$}}
\newcommand{\bj} { \mbox{\boldmath $j$}}
\newcommand{\bk} { \bm{k} }
\newcommand{\bl} { \mbox{\boldmath $l$}} 
\newcommand{\bmm} { \mbox{\boldmath $m$}}
\newcommand{\bn} { \mbox{\boldmath $n$}}
\newcommand{\bo} { \mbox{\boldmath $o$}}
\newcommand{\bp} { \bm{p} }
\newcommand{\bq} { \bm{q} }
\newcommand{\br} { \boldsymbol{r}}
\newcommand{\bs} { \mbox{\boldmath $s$}}
\newcommand{\bt} {\boldsymbol{t}} 
\newcommand{\bu} { \mbox{\boldmath $u$}}
\newcommand{\bv} { \mbox{\boldmath $v$}}
\newcommand{\bw} { \mbox{\boldmath $w$}}
\newcommand{\bx} { \mbox{\boldmath $x$}}
\newcommand{\by} { \mbox{\boldmath $y$}}
\newcommand{\bz} { \mbox{\boldmath $z$}}
\newcommand{\bA} { \mbox{\boldmath $A$}}
\newcommand{\bB} { \mbox{\boldmath $B$}}
\newcommand{\bC} { \mbox{\boldmath $C$}}
\newcommand{\bD} { \mbox{\boldmath $D$}}
\newcommand{\bF} { \mbox{\boldmath $F$}}
\newcommand{\bG} { \mbox{\boldmath $G$}}
\newcommand{\bH} { \mbox{\boldmath $H$}}
\newcommand{\bI} { \mbox{\boldmath $I$}}
\newcommand{\bJ} { \mbox{\boldmath $J$}}
\newcommand{\bK} { \mbox{\boldmath $K$}}
\newcommand{\bL} { \mbox{\boldmath $L$}}
\newcommand{\bM} { \mbox{\boldmath $M$}}
\newcommand{\bN} { \mbox{\boldmath $N$}}
\newcommand{\bO} { \mbox{\boldmath $O$}}
\newcommand{\bP} { \mbox{\boldmath $P$}}
\newcommand{\bQ} { \boldsymbol{Q} }
\newcommand{\bR} { {\mathbf R} }
\newcommand{\bS} { \mbox{\boldmath $S$}}
\newcommand{\bT} { \mbox{\boldmath $T$}}
\newcommand{\bU} { \mbox{\boldmath $U$}}
\newcommand{\bV} { \mbox{\boldmath $V$}}
\newcommand{\bW} { \mbox{\boldmath $W$}}
\newcommand{\bX} { \mbox{\boldmath $X$}}
\newcommand{\bY} { \mbox{\boldmath $Y$}}
\newcommand{\bZ} { \mbox{\boldmath $Z$}}
\newcommand{\bzero} { \mbox{\boldmath $0$}}
\newcommand{\bfell} {\mbox{\boldmath $ \ell $}}

%
%
\newcommand{\dou}{\partial}
\newcommand{\leftjb} {[\![}
\newcommand{\rightjb} {]\!]}
\newcommand{\ju}[1]{ \leftjb #1 \rightjb }
\newcommand{\D}[1]{\mbox{d}{#1}} 
\newcommand{\grad}{\mbox{\boldmath $\nabla$}}
\newcommand{\modulus}[1]{|#1|}
\renewcommand{\div}[1]{\grad \cdot #1}
\newcommand{\curl}[1]{\grad \times #1}
\newcommand{\mean}[1]{\langle #1 \rangle}
\newcommand{\bra}[1]{{\langle #1 |}}
\newcommand{\ket}[1]{| #1 \rangle}
\newcommand{\braket}[2]{\langle #1 | #2 \rangle}
\newcommand{\dbdou}[2]{\frac{\dou #1}{\dou #2}}
\newcommand{\dbdsq}[2]{\frac{\dou^2 #1}{\dou #2^2}}
\newcommand{\Pint}[2]{ P \!\!\!\!\!\!\!\int_{#1}^{#2}}
\newcommand{\Itwo}{{\mathds{1}}}
\newcommand{\Hds}{{\mathds{H}}}
\newcommand{\cH}{{\cal H}}
\newcommand{\cS}{{\cal S}}

%
%
\newcommand{\prn}[1] {(\ref{#1})}
\newcommand{\sect}[1] {Section~\ref{#1}}
\newcommand{\Sect}[1] {Section~\ref{#1}}

%
%
\newcommand{\uncon}[1]{\centerline{\epsfysize=#1 \epsfbox{/usr2/yogeshwar/styles/construction.pdf}}}
\newcommand{\checkup}[1]{{(\tt #1)}\typeout{#1}}
\newcommand{\ttd}[1]{{\color[rgb]{1,0,0}{\bf #1}}}
\newcommand{\red}[1]{{\color[rgb]{1,0,0}{\protect{#1}}}}
\newcommand{\blue}[1]{{\color[rgb]{0,0,1}{#1}}}
\newcommand{\green}[1]{{\color[rgb]{0.0,0.5,0.0}{#1}}}
\newcommand{\citebyname}[1]{\citeauthor{#1}\cite{#1}}
\newcommand{\myfigwidth}{0.95\columnwidth}
\newcommand{\myhalffig}{0.475\columnwidth}
\newcommand{\mythirdfig}{0.33\columnwidth}
\newcommand{\signum}[0]{\mathop{\mathrm{sign}}}
\newcommand{\skup}{\ket{s \uparrow}}
\newcommand{\skdn}{\ket{s \downarrow}}
\newcommand{\pkup}{\ket{p \uparrow}}
\newcommand{\pkdn}{\ket{p \downarrow}}
\newcommand{\sbup}{\bra{s \uparrow}}
\newcommand{\sbdn}{\bra{s \downarrow}}
\newcommand{\pbup}{\bra{p \uparrow}}
\newcommand{\pbdn}{\bra{p \downarrow}}

\newcommand{\Eqn}[1] {Eqn.~(\ref{#1})}
\newcommand{\Fig}[1]{Fig.~\ref{#1}}

\title{Initial State Dependent Dynamics Across Many-body Localization Transition}
\author{Yogeshwar Prasad}
\affiliation{Theory Division, Saha Institute of Nuclear Physics, 1/AF Bidhannagar, Kolkata 700 064, India}
\author{Arti Garg}
\affiliation{Theory Division, Saha Institute of Nuclear Physics, 1/AF Bidhannagar, Kolkata 700 064, India}
 \affiliation{Homi Bhabha National Institute, Training School Complex, Anushaktinagar, Mumbai 400094, India}
\vspace{0.2cm}
\begin{abstract}
\vspace{0.3cm}
       {We investigate quench dynamics across many-body localization (MBL) transition in an interacting one dimensional system of spinless fermions with aperiodic potential. We consider a large number of initial states characterized by the number of kinks, $N_{kinks}$, in the density profile, such that equal number of sites are occupied between any two consecutive kinks.
         We show that on the delocalized side of the MBL transition the dynamics becomes faster with increase in $N_{kinks}$ such that the decay exponent, $\gamma$, in the density imbalance increases with increase in $N_{kinks}$. The growth exponent of the mean square displacement which shows a power-law behaviour $\langle x^2(t) \rangle \sim t^\beta$ in the long time limit is much larger than the exponent $\gamma$ for 1-kink and other low kink states though $\beta \sim 2\gamma$ for a charge density wave state. As the disorder strength increases $\gamma_{N_{kink}} \rightarrow 0$ at some critical disorder, $h_{N_{kinks}}$ which is a monotonically increasing function of $N_{kinks}$. A 1-kink state always underestimates the value of disorder at which the MBL transition takes place but $h_{1-kink}$ coincides with the onset of the sub-diffusive phase preceding the MBL phase. This is consistent with the dynamics of interface broadening for the 1-kink state. We show that the bipartite entanglement entropy has a logarithmic growth $a \ln(Vt)$ not only in the MBL phase but also in the delocalised phase and in both the phases the coefficient $a$ increases with $N_{kinks}$ as well as with the interaction strength $V$. We explain this dependence of dynamics on the number of kinks in terms of the normalized participation ratio of initial states in the eigenbasis of the interacting Hamiltonian.} 
         
\vspace{0.cm}
\end{abstract} 
\maketitle
\section{I. Introduction}
Interplay of disorder and interactions result in exotic phenomena. Many-body localization (MBL) is one such phenomenon where Anderson localization~\cite{Anderson}  persists even in the presence of interactions, at least for certain range of interactions~\cite{Basko,Gornyi,Huse_rev,Abanin_rev,Abanin2,Alet_rev,Ehud_rev}. Theoretically MBL has been proved to exist in 1-dimensional systems with short range interactions~\cite{Imbrie}. MBL to delocalization transition is associated with a transition from a non-ergodic to ergodic phase and hence can be characterized by statistics of level-spacing of the many-body eigen spectrum ~\cite{Mehta1990,Huse2007,Alet} and eigenstate thermalization hypothesis~\cite{Deutsch,Srednicki,Rigol}. Though the localized nature of many-body eigenstates is identified using the statistics of many-body eigen-functions in the Fock space~\cite{Alet_rev,Serbyn,Santos,Luitz2020,Tikhonov2016}, scaling of subsystem entanglement entropy~\cite{Alet_rev,Alet,Huse2013,Bardarson,Sdsarma,garg}, and scaling of local density of states and scattering rates~\cite{Atanu}. 

MBL systems have strong memory of initial states which is a reflection of their non-ergodic nature. Starting from any initial state the system in the MBL phase carries strong signatures of it even at very long time. Therefore, long time dynamics of the density imbalance starting from a charge density wave (CDW) state has been used extensively, both experimentally~\cite{expt,Luschen2017b,Kohlert2019} and computationally~\cite{Mirlin_long,Mirlin_AA,Mirlin_imb_HF,Imbalance_Knap,Poitr_imb} to track the MBL transtion. On the delocalized side of the MBL transition, the density imbalance decays to zero in the long time limit because the system looses memory of the initial state while in the MBL phase the imbalance saturates to a finite value in the long time limit. Furthermore, in the delocalized phase the density imbalance shows a power-law decay $I(t) \sim t^{-\gamma}$~\cite{Luschen2017b,Kohlert2019,Yevgeny_rev,Mirlin_long,Mirlin_AA,Mirlin_imb_HF,Mirlin_imb_del} after the initial rapid decay. As the disorder strength increases the decay exponent decreases and at the MBL transition point $\gamma \rightarrow 0$~\cite{Luschen2017b,Kohlert2019}. In systems with random disorder, a large regime of the delocalized phase has a subdiffusive dynamics (with $\gamma < 1/2$) preceding the MBL phase~\cite{Luschen2017b,Kohlert2019,Mirlin_AA,Mirlin_long,Mirlin_imb_HF,Imbalance_Knap} which is associated with the presence of rare-extremely localized regions in otherwise delocalized phase (Griffiths effects)~\cite{Gopalak,Agarwalk}. Interestingly, the slow subdiffusive dynamics has also been seen in systems with quasi-periodic potential~\cite{Luschen2017b,Kohlert2019,Bera,Mirlin_AA,Prasad2021} but there is no consensus on the mechanism behind slow dynamics in these deterministic systems.

Thus, quench dynamics has played a crucial role in understanding the delocalized side of the MBL transition. It has also raised some subtle issues about the MBL transition point and the stability of the MBL phase. Imbalance calculation for large size chains have shown that the decay exponent remains non zero for much larger values of disorder strength beyond the transition point known from other criterion like level spacing ratio which are generally obtained from exact diagonalization for smaller systems~\cite{Mirlin_long,Poitr}. But surprisingly almost all the computational and experimental works in this direction have focused on CDW as the initial state. Recently, in an experiment on 2-dimensional bosons the density imbalance was studied starting from an initial state in which all the particles are confined to one half of the system~\cite{Choi2016_Science} followed up by theoretical works on similar initial state for 1-dimensional models ~\cite{Mirlin_imb_HF,Imbalance_Knap,Pollmann_DWmelt,Modak2022}, but a systematic quench analysis for many different initial states has not been performed in detail in most of the earlier works barring a few exceptions~\cite{Guo_expt_ME,ChandaME}. Hence, many interesting questions like how the dynamical exponent and the critical disorder at which $\gamma \sim 0$ depends on the initial state have remained unanswered. 
Previous numerical works have also indicated that the exponent from the density imbalance obeys a simple relation with the exponent obtained from the time evolution of the mean square displacement of a density fluctuation obtained from the time dependent density-density correlation function~\cite{Yevgeny_rev,Luitz2020}. But the quench dynamics depends upon the initial state in which the system is prepared while the density-density correlation function is obtained from an infinite temperature ensemble average, and hence is independent of the initial state. A natural question that arises is how the decay exponents from the imbalance starting from various initial states are related to the exponent obtained from an initial state independent mean square displacement? These are some important questions which have been addressed in this work. 

With this motivation, we study quench dynamics across MBL transition starting from a large number of matrix product initial states which are characterized by the number of kinks in the density profile of the chain such that equal number of sites are occupied between any two consecutive kinks. A schematic of initial states with different number of kinks $N_{kinks}$ is shown in \Fig{kinks}. 1-kink state has all the particles on one half of the chain while the CDW state has $N_{kinks}=L-1$ kinks in it. We study time evolution of the corresponding density imbalance and the sublattice entanglement entropy for various initial states across the MBL transition.  We also calculate the mean square displacement $\langle x^2(t)\rangle$ from the density-density correlation function and compare the decay exponent from the quench dynamics of various initial states with the growth exponent of $\mean{x^2(t)}$ on the ergodic side of the MBL transition point. To be specific, we study quench dynamics in a system of spinless fermions in one-dimension in the presence of a deterministic aperiodic potential and nearest neighbour interactions. This model has been studied before in detail in context of MBL~\cite{Sdsarma,Subroto,garg,garg_lr,Prasad2021} but the quench dynamics in the presence of nearest neighbour interactions has not been explored yet even for a CDW initial state. Below we summarise the main results from this work.

\begin{figure*}[ht]
  \begin{center}
    \vskip0.5cm
    \hspace{-1cm}
  \includegraphics[width=6.6in]
                  {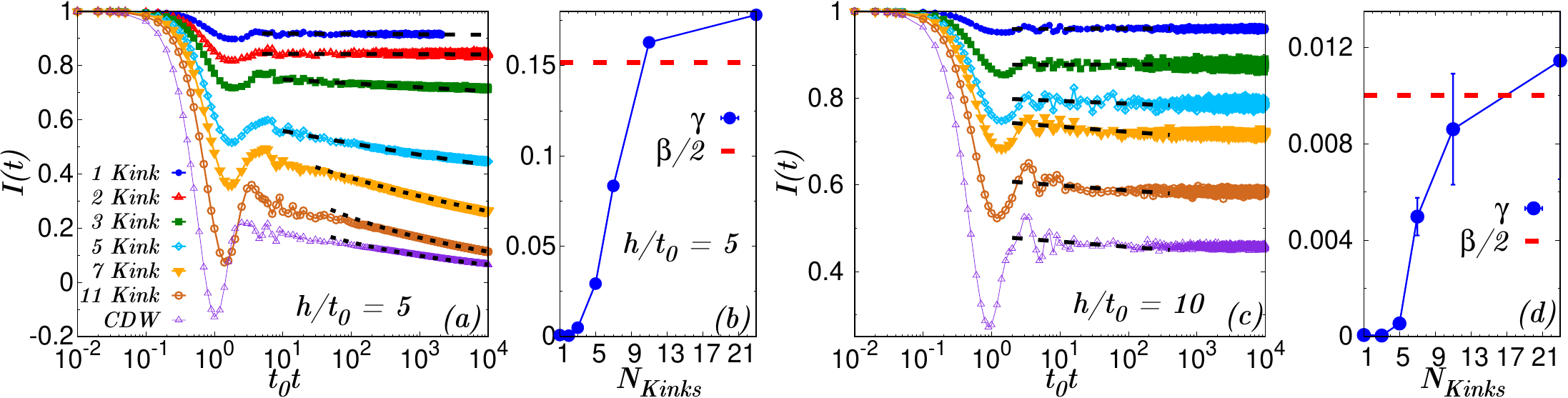}
  \caption{The density imbalance $I(t)$ as a function
           of time $t$ for $h = 5t_0$ and $10t_0$ at $V = t_0$ and $L=24$ for various kink initial
           states. The MBL transition point for the model in \Eqn{Ham} is $h_c=6.3t_0$ at $V=t_0$ (\Fig{rn}). Dashed lines show the power-law fit to the form $t^{-\gamma}$. Panel
           $(b,d)$ show the imbalance decay exponent $\gamma$ obtained from power-law fits
           for $h = 5t_0$ and $10t_0$ respectively. Dashed red-line in these panels  is the
           exponent $\beta/2$ obtained from mean square displacement. Note that $\gamma \sim 0$ for all the initial states deep in the MBL phase though one can still see $\gamma$ increasing monotonically with $N_{kinks}$.}
  \label{Imb_kinks}
\vskip-1cm
\end{center}
\end{figure*}

\begin{itemize}
  \item On the delocalised side of the MBL transition point where the imbalance has a power-law decay, $I(t) \sim t^{-\gamma}$, in the intermediate to long time regime, $\gamma$ increases monotonically with $N_{kinks}$ being maximum for the CDW state. In the MBL phase the imbalance saturates after initial time decay for all the kink states but the saturation value $I_{sat}$ decreases as $N_{kinks}$ increase, being minimum for the CDW state (\Fig{Imb_kinks}). We explain this trend of dynamics in terms of the normalised participation ratio of the initial state in the eigenbasis of the Hamiltonian under consideration.

\begin{figure}[ht]
  \begin{center}
    \vskip0.5cm
  \includegraphics[width=3.3in]
                  {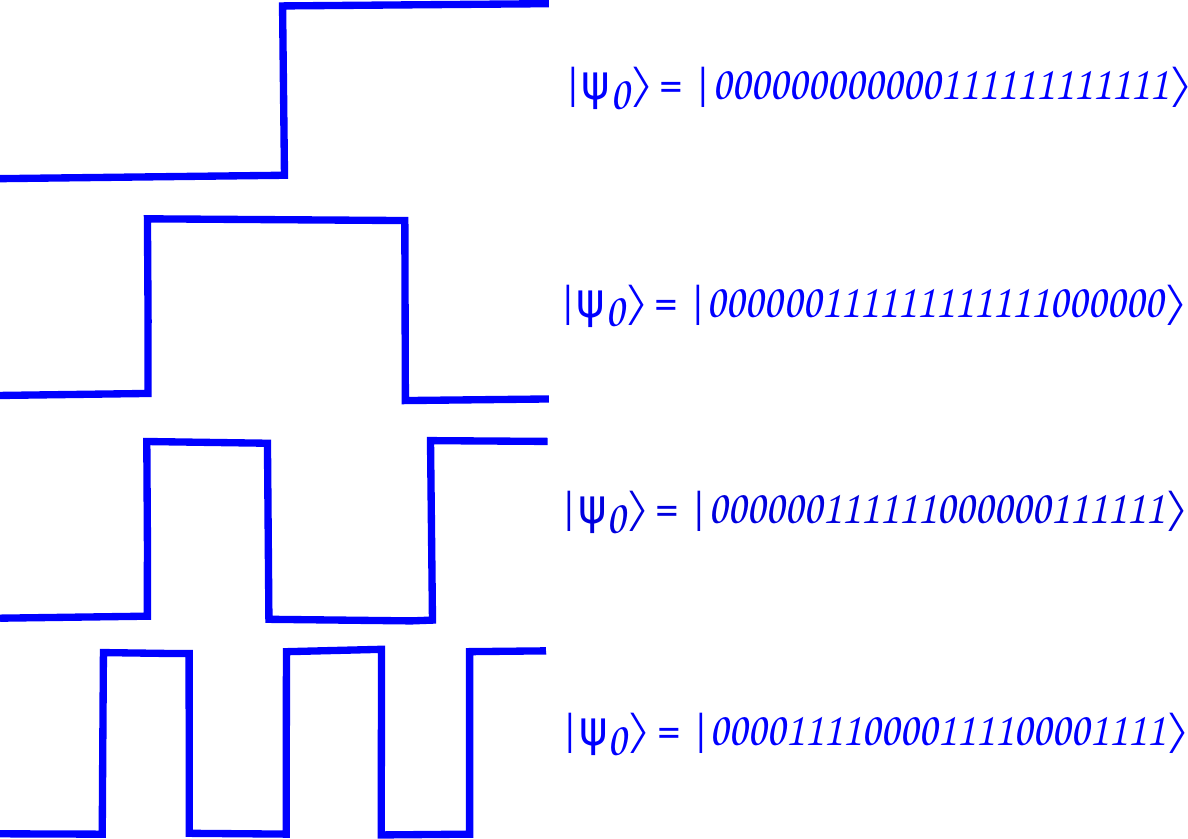}
  \caption{For a half-filled system of spin-less fermions, we study various initial states characterized by the number of kinks in the density profile such that equal number of sites are occupied between any two consecutive kinks. Schematic diagram of initial states with $1, 2, 3$ and $5$ kinks for $L = 24$ sites chain is shown here.}
  \label{kinks}
\vskip-1cm
\end{center}
\end{figure}

\item Generally delocalization to MBL transition point is identified as the disorder strength at which $\gamma \rightarrow 0$ coming from the delocalized side~\cite{Luschen2017b,Kohlert2019,Mirlin_AA,Mirlin_long,Mirlin_imb_HF}. We show that  for a $N_{kinks}$ initial state $\gamma$ goes to zero at some critical field, $h_{N_{kinks}}$, such that $h_{N_{kinks}}$ is a monotonically increasing function of $N_{kinks}$ (\Fig{diffh}). Thus, as the disorder strength increases, $\gamma\rightarrow 0$ first for the 1-kink state such that $h_{1kink} < h_c$ and the imbalance from large kink states (e.g., a CDW state) continues to show a power-law decay with a finite $\gamma$ for a much larger value of disorder with $h_{CDW} > h_c$. Here, $h_c$ is the transition point determined from level spacing ratio of eigen-energies. 
  
\item The dynamics from the time evolution of mean square displacement $\langle x^2 \rangle$, obtained from the time dependent density-density correlation function, is much faster than that from the density imbalance of low-kink states. $\mean{x^2(t)}$ shows a power-law growth in the long time limit, $\langle x^2(t) \rangle \sim t^{\beta}$, with $\beta \gg \gamma_{1kink},\gamma_{3kinks}$ though $\beta/2$ is close to $\gamma$ from initial state with large number of kinks ( \Fig{gamma2}).
  
  \item We also study the melting of interface as an alternate probe of dynamics for 1-kink state. The melting dynamics of the interface is completely consistent with the time evolution of the density imbalance for 1-kink state and is much slower than the dynamics of the CDW state (\Fig{xt}). Thus, a quench dynamics study starting from a 1-kink or other low kink state will always underestimate the critical value of disorder required to cause many-body localization. In fact, 1-kink and other low kink states only indicate the onset of localization of a finite fraction of many-body states and hence the subdiffusive phase which appears due to multifractal nature of eigenstates close to the MBL transition.
\item The sublattice entanglement entropy shows a logarithmic growth $S(t) \sim a \ln(Vt)$ after initial rapid growth, both, for the ergodic phase as well as the MBL phase. The coefficient of the $\ln(Vt)$ term not only increases monotonically with the number of kinks being maximum for the CDW state (\Fig{SKinks}), but also increases significantly with the interaction strength $V$ (\Fig{EEV}) indicating that the dependence on $V$ is faster than $\ln(Vt)$. The coefficient of $\ln(Vt)$ term inside the MBL phase is vanishingly small. All these observations put a question mark on earlier explanations of the logarithmic growth of entanglement entropy in terms of the local integrals of motion which exist only in the MBL phase~\cite{Serbyn}.
\end{itemize}
The rest of the paper is organized as follows. In Section II, we introduce the model explored in this work. In section III,  we describe the dynamics from time evolution of the density imbalance for various kink initial states. We also compare the dynamics obtained from imbalance with that from time dependent mean square displacement which is the second moment of the density-density correlation function. In section IV, we study the melting of the interface for 1-kink state and show that it is consistent with imbalance for 1-kink state having a dynamics much slower than that of a CDW state. In section V we discuss the growth of sublattice entanglement entropy starting from various kink initial states. Finally we summarize our results and conclude with some remarks and open questions.

\section{II. Model}
We consider a 1D model of spinless fermions in the presence of an aperiodic potential and nearest neighbour interactions described by the
Hamiltonian :
\begin{eqnarray}
  \label{eqn:Hamilt}
 H=-t_0\sum_{\langle i,j \rangle}(c^\dagger_i c_j + h.c.)+ \sum_i h_i n_i \nonumber \\
 + V \sum_i n_i n_{i+1}.
 \label{Ham}
\end{eqnarray}

Here $h_i = h \cos(2\pi \alpha i^n + \phi)$ represents a deterministic aperiodic potential with strength $h$, $\alpha$ is an irrational number which we chose to be  $(\frac{\sqrt{5} - 1}{2})$, $\phi\in[0,2\pi)$ is a random phase taken from a uniform distribution and $n$ is a real number. $t_0$ is the strength of nearest neighbour hopping amplitude and $V$ is the strength of nearest neighbour repulsion between fermions. We study this model at half-filling with open boundary conditions.

For the non-interacting model ($V=0$), all the single-particle states are localized for any value of $n$ for $h >2t_0$. For $n < 1$, the system shows single particle mobility edges at $E_c = \pm |2t_0-h|$ for $h/t_0 < 2$~\cite{Fishman,Sarma1990,Sarma_nonint} while for $n=1$ $h(i)$ gives the quasiperiodic Aubry-Andre potential~\cite{AA}. We chose to work with $n=0.5$ for which all the many-body eigenstates of the non-interacting half-filled system are delocalised for $h < 2t_0$ ~\cite{Subroto,garg} while for $h > 2t_0$ all the many-body eigenstates of the non-interacting system are localized. To obtain the critical disorder $h_c$ at which delocalization to MBL transition takes place in the presence of interactions, we calculated the average level spacing ratio for several system sizes. The critical disorder from the data collapse is $h_c \sim 6.3t_0$ for $V=t_0$. Details are given in Appendix A.

Though the interacting model in \Eqn{Ham} has been studied before in the context of MBL~\cite{Subroto,garg,garg_lr,Prasad2021} but the quench dynamics and the dynamics from the mean square displacement has not been explored yet for this model. In the following sections we discuss the quench dynamics across the MBL transition in this model starting from various kink initial states shown schematically in \Fig{kinks} and show its comparison with the time evolution of density-density correlation function which is calculated in the limit of infinite temperature ensemble average. Most of the result presented below are for $V=t_0=1$ unless specified.
\section{III. Density Imbalance for various kink initial states}
We study dynamics of the system after a quench starting from various kink initial states which are schematically shown in \Fig{kinks}. For 1-kink state all the particles are located in the first half of the chain such that $|\Psi_0\rangle = \prod_{i=0}^{L/2-1}c^\dagger_i|0\ra$. For 2-kink states all the particles are distributed in the middle of the lattice leaving equal number of empty sites on both sides of the kinks such that $|\Psi_0\ra = \prod_{i=L/4-1}^{3L/4}c^\dagger_i|0\ra$. For higher kinks states with $N_{kinks}$ number of kinks in the density profile, equal number of particles are distributed between any two consecutive kinks. The CDW state has $N_{kinks}=L-1$ for a $L$ site chain with $|\Psi_0\ra =\prod_{i=0}^{L/2-1} c^\dagger_{2i}|0\ra$. The particle-hole symmetric counterparts of these states have the same dynamics as these states.

The corresponding density imbalance for the half-filled system is defined as
\be
I(t) = \frac{2}{L} \biggl [\sum_{i_1} \mean{n_{i_1}(t)} - \sum_{i_0} \mean{n_{i_0}(t)}\biggr ]
\label{I_kinks}
\ee
where $i_1$ represent the occupied sites at $t=0$ and $i_0$ are the unoccupied sites at $t=0$ for a particular initial state.
Starting from  $|\Psi_0\ra$, we let the state evolve w.r.t the Hamiltonian in \Eqn{Ham} to obtain the time evolved state $|\Psi(t)\ra=exp(-iHt)|\psi_0 \ra$ and calculate $I(t)$ as a function of time which is then averaged over many independent disorder realizations. Time evolution is carried out numerically using Chebychev polynomial method~\cite{Weiss,Fehske,Holzner,Halimeh,Soumya}. The results presented below are for $L=24$ sites chain and disorder averaging was done over 150 independent configurations.

After the initial rapid decay, $I(t)$ follows a power-law decay $I(t) \sim t^{-\gamma}$ for intermediate to large time which has also been observed in various previous works~\cite{Yevgeny_rev,Mirlin_AA,Mirlin_long}.  The power-law decay in the delocalized phase can be explained in terms of the mixing of slow and fast modes~\cite{Mirlin_imb_del}. The exponent $\gamma$ has been used as a measure of the nature of transport; with $\gamma = 1/2$ for a  diffusive system,  $\gamma=1$ for a ballistic system while $\gamma=0$ for a localized system.
\begin{figure}[ht]
  \begin{center}
    \vskip0.5cm
    \hspace{-1cm}
  \includegraphics[width=3.3in]
                  {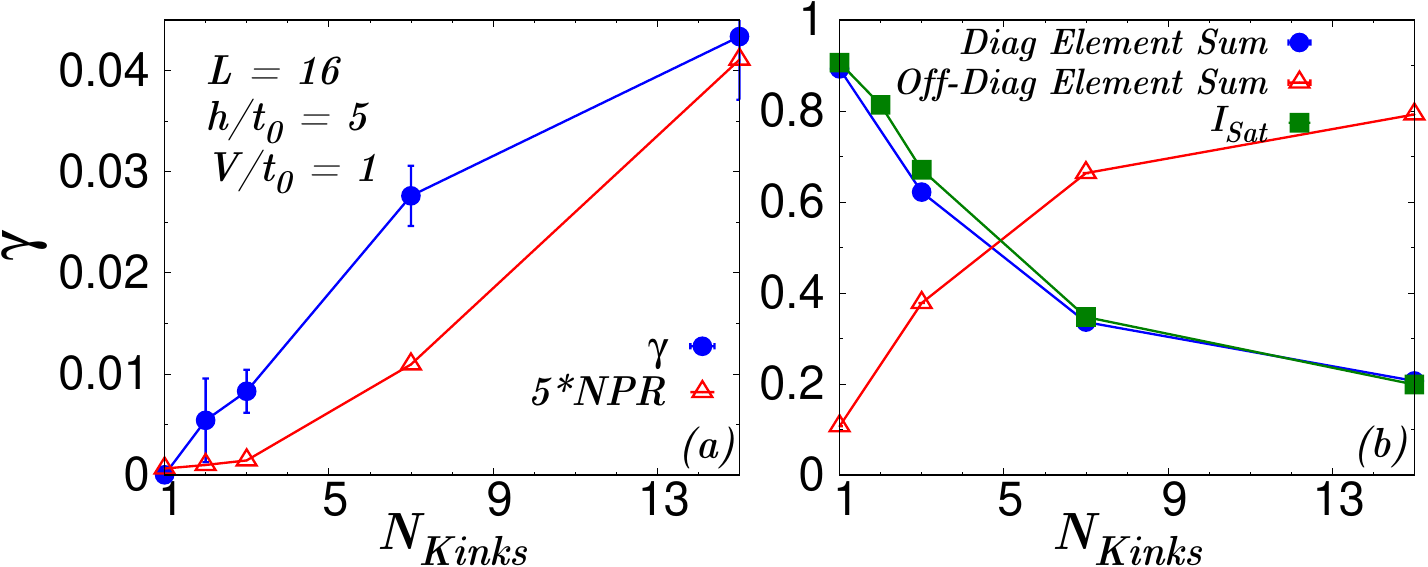}
  \caption{Panel $(a)$ shows the comparison of $\gamma$ with $NPR$
           at $h=5t_0, V=t_0$ for $L=16$. Like $\gamma$, $NPR$ also
           increases monotonically as we go from 1-kink state to higher
           kink states, being maximum for the CDW state indicating that
           the CDW state gets contribution from a larger fraction of eigenstates. In panel $(b)$ we show $I_{sat}$ along with the diagonal and off-diagonal
           elements of $I(t=0)$ for various kink initial states
           at $h=5t_0, V=t_0$ for $L=16$.}
  \label{L16}
\vskip-1cm
\end{center}
\end{figure}

\Fig{Imb_kinks} shows the density imbalance $I(t)$ on two sides of the MBL transition point $h_c$, namely, for $h=5t_0$ and $h=10t_0$. As shown in \Fig{Imb_kinks}, even on the delocalized side of the MBL transition point, the dynamics is not the same for all the initial states. For initial states with 1-kink and 2-kink states, after initial rapid decay, the imbalance does not show any power-law decay such that $\gamma\sim 0$ and the imbalance saturates in the long time limit. Interestingly, $\gamma$ increases monotonically  with $N_{kinks}$ in the initial states, being maximum for the CDW initial state. This indicates that the system relaxes faster if prepared in initial states with larger number of kinks. Deep in the MBL phase, that is for  $h=10t_0, V=t_0$, the imbalance does not show any significant decay after the initial rapid decay and $\gamma \le 0.01$ for all the kinks. But the saturation value of the imbalance $I_{sat}$ decreases as $N_{kinks}$ increases, being largest for the 1-kink state and minimum for the CDW state again indicating faster relaxation of the system for the CDW state compared to initial states with less number of kinks. Interestingly, time evolution of the density imbalance in the MBL phase is very similar to that for an Anderson localized phase of the corresponding non-interacting system though the dynamics in the delocalized phase of the interacting and non-interacting system are very different. In the non-interacting case, there is no power-law decay for any initial state even in the delocalized phase as shown in Appendix B.
\begin{figure*}[ht]
  \begin{center}
    \vskip0.5cm
    \hspace{-1cm}
  \includegraphics[width=6.3in]
                  {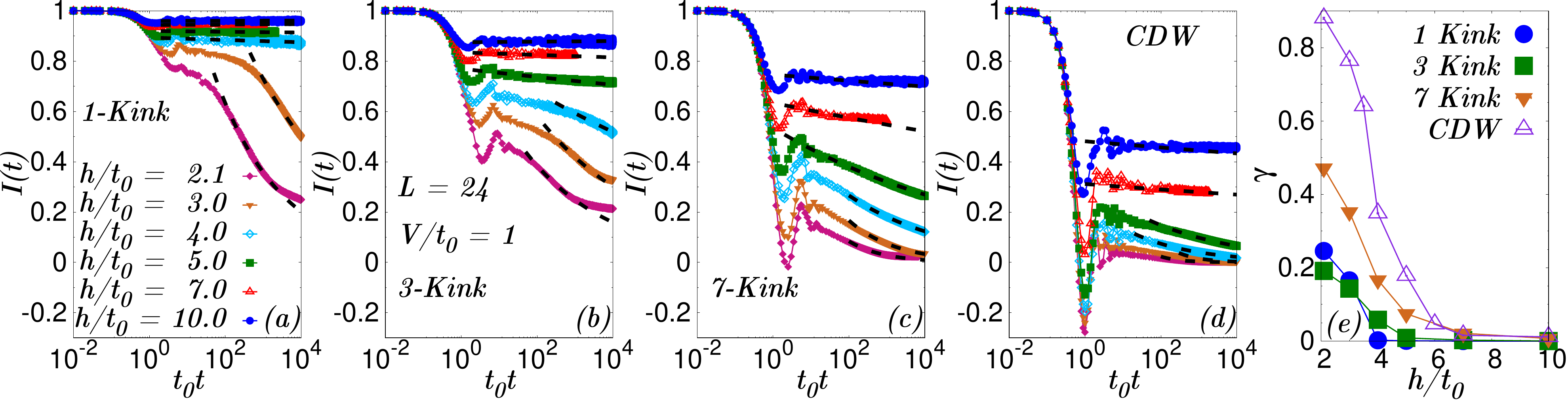}
  \caption{The density imbalance $I(t)$ as a function of time $t$
           for $1, 3, 7$ kink and CDW initial states for various disorder strengths
           at $V=t_0$ for $L=24$. Panel $(e)$ shows $\gamma$ as a function
           of disorder strength $h/t_0$ for various initial states. Starting from a 1-kink initial state $I(t)$ 
           shows saturation after the initial rapid decay for $h \ge h_{1kink} =4t_0 $ which is less than $h_c=6.3t_0$ obtained from level spacing ratio(\Fig{rn}),
           while for a higher kink state, such as 3-kink or 7-kink initial state,
           imbalance shows power-law decay upto much larger values of the aperiodic potential.}
  \label{diffh}
\vskip-1cm
\end{center}
\end{figure*}

In contrast to this, deep in the delocalized phase (e.g. $h=V=t_0$) where all the many-body eigenstates are extended, initial states with less number of kinks show slower decay of the density imbalance compared to initial states with large $N_{kinks}$ only for a short time which can be explained in terms of the lower connectivity of low kink states in the Fock space compared to larger kink states for a system with nearest neighbour hopping. But eventually in the long time limit the decay rate increases for all the initial states such that imbalance goes to zero as it should for an ergodic system. The long time imbalance seems to decay faster than power-law, though in the limit of long time, power-law fit seems to work well with decay exponent $\gamma \ge 1$ for some of the initial states. It indicates the ballistic or super-ballistic transport, which is a reminiscent of the ballistic transport in the corresponding non-interacting models with deterministic potentials. The long time decay rates do not show a systematic trend as a function of number of kinks in the initial state deep inside the delocalized phase as shown in Appendix C. Therefore, a monotonic trend of $\gamma$ as a function of $N_{kinks}$ is a good indicator of the fact that the system has at least a finite fraction of the  many-body states localized.

Above results indicate that as long as the system has a finite fraction of many-body states localized, we see a systematic dependence of the long-time dynamics on the number of kinks in initial states, either in terms of the kink dependence of $\gamma$ or $I_{sat}$.  
We explain these results in the following way. Any initial state $|\Psi_0\rangle$, with $N_{kinks}$ number of kinks, can be written as a linear combination of the many-body eigenstates of the Hamiltonian in \Eqn{Ham} as $|\Psi_0\rangle=\sum_n C_n(N_{kink}) |\Phi_n\rangle$ where $H|\Phi_n\rangle=E_n|\Phi_n\rangle$. We estimate the fraction of eigenstates that contribute to a given
initial state through the calculation of normalized participation ratio ($NPR$)
\be
NPR(N_{kink}) = \frac{1}{N}\frac{1}{\sum_{n=1}^N|C_n(N_{kink})|^4}
\label{NPR_kink}
\ee
where $N$ is the dimension of the Fock space. Note that NPR calculation requires exact diagonalization of the Hamiltonian in \Eqn{Ham}, and hence we have shown results for $L=16$ though imbalance in all earlier plots has been calculated for $L=24$. 
\Fig{L16} shows $NPR$ vs  $N_{kinks}$ for $h=5t_0, V=t_0$ which has been averaged over many independent disorder configurations. $NPR$ is vanishingly small for 1-kink state and it increases monotonically as we go from 1-kink state to higher kink states, being maximum for the CDW state. This implies that larger fraction of eigenstates contribute to the CDW state as compared to the 1-kink and other lower kink states. Thus, CDW state will have faster time evolution resulting in faster decay of the density imbalance compared to the case of 1-kink or other lower kink states.
This is clearly visible in \Fig{L16} where we have shown a comparison of $\gamma$ and $NPR$ both calculated for $L=16$ and $h=5t_0, V=t_0$. Deep in the MBL phase, lower values of $NPR$ for lower kink initial states, again imply slower decay rate of these initial states for initial time resulting in larger saturation values of the density imbalance $I_{sat}$ for lower kink states.

Using the expansion of the initial state in terms of the eigenstates of Hamiltonian in \Eqn{Ham}, we further express the density imbalance as
\be
I(t)=\sum_{n,m} C_n C_m e^{-it(E_m-E_n)} \langle \Phi_n|\hat{I}|\Phi_m\rangle
\ee
This can be written as the sum of diagonal and off-diagonal terms as follows
\bea
I(t) = \sum_n|C_n|^2 M_{nn} + \sum_{n\ne m} C_nC_m e^{-it(E_m-E_n)} M_{nm} \nonumber \\
= I_{diag} + I_{off-diag}(t)
\label{imb_en}
\eea

where $\hat{I}$ is the operator corresponding to the density imbalance for the corresponding initial state and $M_{nm} = \langle \Phi_n|\hat{I}|\Phi_m\rangle$. The decay in imbalance happens only through the off-diagonal elements and the saturation value of the density imbalance in the long time limit is given by $I_{diag}$ provided  there are no degeneracies in the eigenspectrum.
The right panel of \Fig{L16} also shows a comparison of the long time value of imbalance $I_{sat}$ with the diagonal element $I_{diag}$. We see that $I_{sat} \sim I_{diag}$ indicating that $I_{diag}$ really provides a good estimate of the saturation value.
\begin{figure*}[ht]
  \begin{center}
    \hspace{-0.8cm}
  \includegraphics[width=5.3in,angle=0]
                  {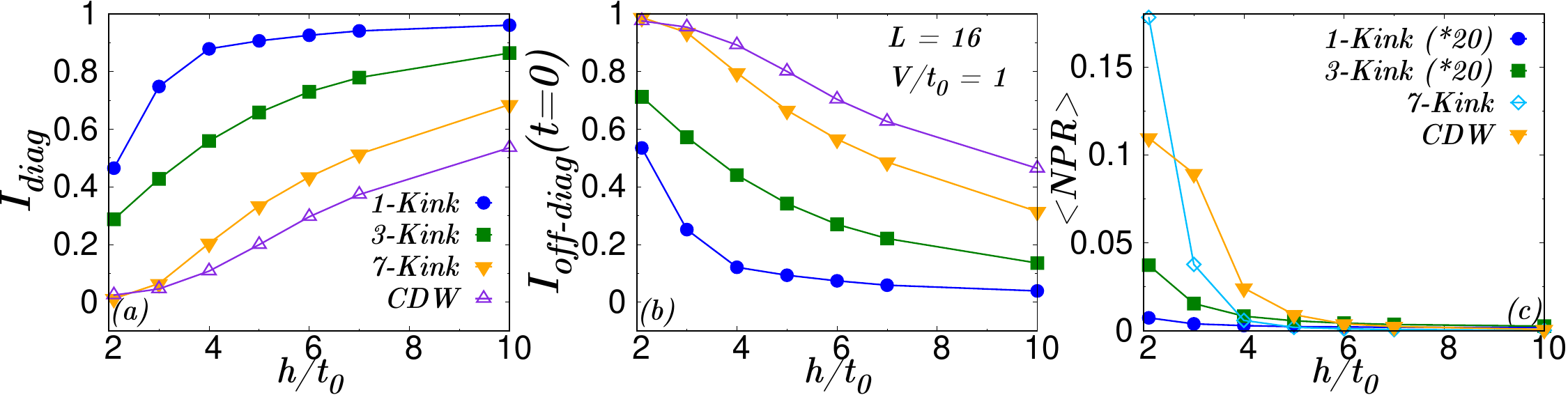}        
  \caption{{Panel (a,b) show $I_{diag}$  and $I_{off-diag}(t=0)$ as a function of
           the disorder strength $h/t_0$ for various kink initial states for $L=16$.}
           As the strength of aperiodic potential increases, $I_{diag}$ approaches $1$ first for lower kink states while the higher kink initial states require much larger $h$ to get $I_{diag}\rightarrow 1$. Panel (c) shows $NPR$ of various initial states in the eigenbasis of the Hamiltonian vs $h$.}
  \label{matel}
\vskip-1cm
\end{center}
\end{figure*}

In the right panel of \Fig{L16} we have also shown the off-diagonal elements of $I(t=0)$ for various kink initial states for $h=5t_0,V=t_0$ and $L=16$. Note that $I_{diag}+I_{off-diag}(t=0)=1$ by definition. The diagonal elements decrease as the number of kinks increase, while the off-diagonal elements increase with increase in the number of kinks in the initial states. This also shows that the saturation value of the imbalance, which is close to $I_{diag}$, decreases as the number of kinks $N_{kinks}$ increases.

\subsection{Critical disorder for various kink initial states}
Now we study the disorder dependence of the density imbalance for various kink states. \Fig{diffh} shows $I(t)$ vs time for $V=t_0$ and various disorder strengths with $h > 2t_0$, which is the transition point of the corresponding non-interacting system. As the disorder strength increases, the power-law decay in the density imbalance slowly gets suppressed and  $\gamma$ decreases monotonically with $h$ for all the initial states. Starting from a 1-kink initial state $I(t)$ shows saturation after the initial rapid decay for $h \ge h_{1kink} =4t_0$. Note that $h_{1kink} < h_c=6.3t_0$ obtained from the level spacing ratio. Thus, the density imbalance from 1-kink state underestimates the value of MBL transition point. But $h_{1kink}$ indicates the onset of localization of a fraction of many-body eigenstates or else the imbalance would have decayed to zero. Furthermore, the density imbalance starting from higher kink states, such as 7-kink or CDW initial state,  shows power-law decay with non-zero value of $\gamma$ for much larger values of the aperiodic potential. The critical disorder $h_{N_{kinks}}$ at which the density imbalance of a $N_{kink}$ initial state starts showing a saturation with $\gamma \sim0$ , which is generally considered to be an indication of the onset of MBL phase~\cite{Mirlin_long,Mirlin_imb_HF}, is a monotonically increasing function of $N_{kinks}$.  The transition point obtained from the time evolution of imbalance for large kink states like 7-kink state is close to the one obtained from the level spacing ratio $h_{7kinks} \sim h_c$. But for the CDW state, which has been most extensively used in experimental and theoretical studies $h_{CDW} > h_c$.

The behaviour of $\gamma$ as a function of $h$ for various initial state can be explained in terms of the disorder dependence of NPR of the corresponding initial states in the eigenbasis of the Hamiltonian. Panel [c] of \Fig{matel} shows $NPR$ vs $h$ for $V=t_0$ and $L=16$ for various kink initial states. As the disorder strength increases and a significant fraction of many-body eigenstates get localized, the fraction of eigenstates that contribute to the 1-kink state reduces significantly such that $NPR \sim 10^{-4}$ for $h \ge 4t_0$. This is consistent with $\gamma \rightarrow 0$ for $h \ge 4t_0$ for a 1-kink initial state. In fact as $h$ increases, $NPR$ decreases for all the initial states and the disorder values at which $NPR$ for $N_{kinks}$ state vanishes is close to the $h_{N_{kinks}}$ at which exponent $\gamma \sim 0$. 
We have also shown the trend of the diagonal and off-diagonal elements of $I(t=0)$ as a function of $h$ in \Fig{matel}. In the completely delocalized phase, $I_{diag}=0$ for all the initial states but for any finite fraction of localized many-body eigenstates, $I_{diag}$ is finite and is larger for initial states with lower values of $N_{kinks}$. As the strength of the aperiodic potential increases the diagonal elements approach one indicating suppressed decay of the density imbalance and stronger memory of the initial state.

\begin{figure*}[ht]
  \begin{center}
    \vskip0.5cm
    \hspace{-1cm}
  \includegraphics[width=4.3in]
                  {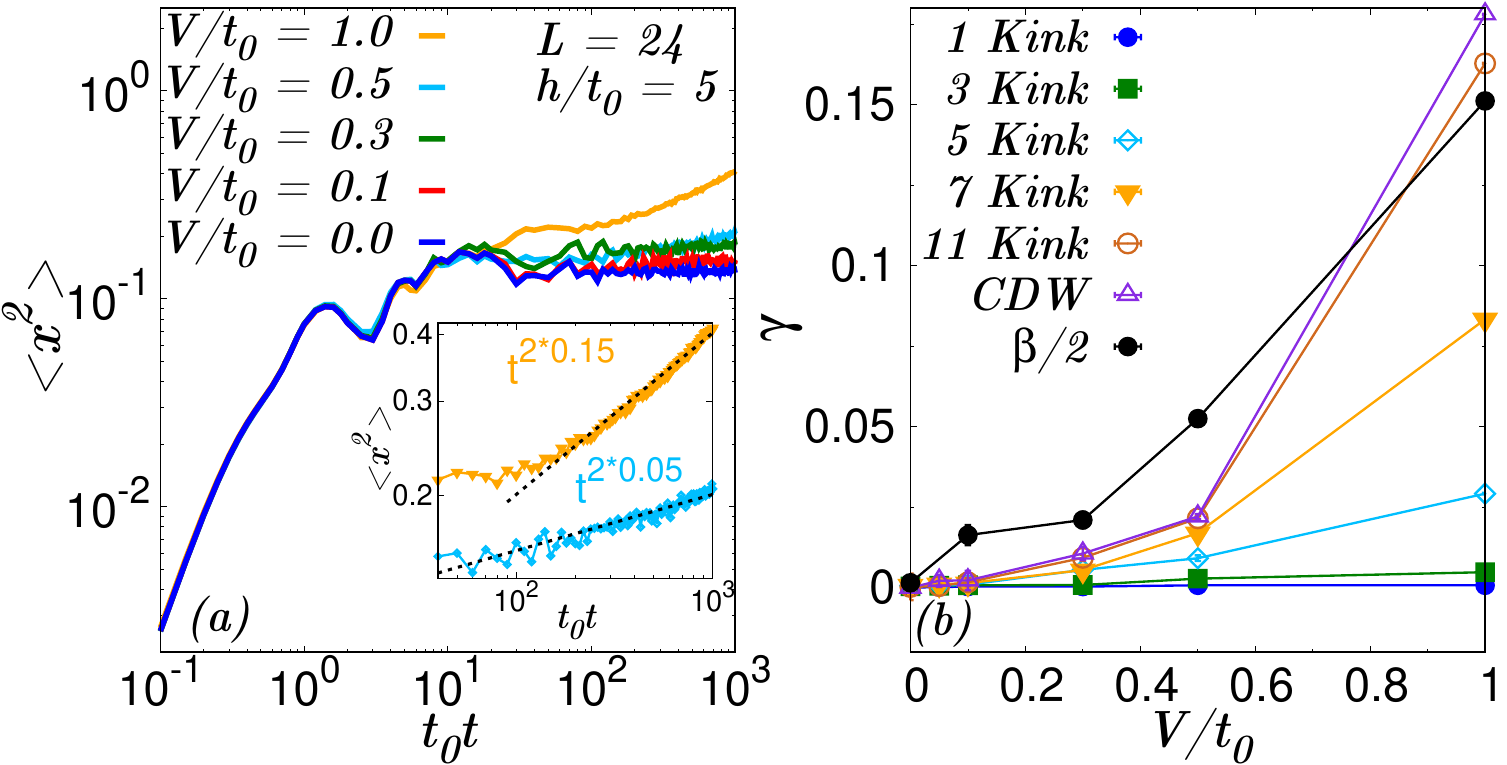}
  \caption{ Panel $(a)$ shows the mean square displacement $\mean{x^2(t)}$ vs $t$ for various values of $V$ at
           $h = 5t_0$. For small values of $V$, $\mean{x^2(t)}$ saturates after the initial rapid growth but
           for larger values of $V$, $\mean{x^2(t)}$ increases as $t^{\beta}$ in the long
           time regime. Panel (b) shows comparison of $\beta/2$ with $\gamma$ obtained from
           the density imbalance for various kink initial states for various values of
           interaction strength $V$ and $h=5t_0$. For any finite value of $V$, $\beta/2 \gg \gamma$ for 1-kink and other lower kink states. $\beta/2$ is closest to $\gamma$ obtained for larger kink states like CDW state.}
  \label{gamma2}
\vskip-1cm
\end{center}
\end{figure*}

There are a few more important observations to be made from this entire analysis. Firstly, even the CDW initial state, which has the fastest dynamics,  shows $\gamma < 1/2$ for intermediate values of $h$ ($ 4t_0\le h<h_c$ for $V=t_0$) indicating a subdiffusive phase preceding the MBL phase. Generally a subdiffusive phase before the MBL transition is associated with rare highly localized regions in otherwise delocalized system~\cite{Yevgeny_rev,Bera2,Varma2016,Mirlin_long,Mirlin_imb_HF,Imbalance_Knap,Agarwalk,Gopalak}. But since the model we are working with has a deterministic potential rather then a random disorder, Griffiths effect can not be the cause for the subdiffusive phase. Interestingly, the disorder value at which the dynamics becomes subdiffusive for the CDW initial state coincides with $h_{1kink}$  where $\gamma \rightarrow 0$ for the 1-kink state. This indicates that the slow dynamics is induced when a significant fraction of many-body states are localized and the extended states of nearby energy are multifractal~\cite{Luitz2020}. For even smaller values of disorder ($2t_0 < h < 4t_0$) a power law fit to the density imbalance starting from a CDW initial state gives a superdiffusive transport with $\gamma > 1/2$. This is a characteristic of generalized Aubry-Andre models where the non-interacting delocalized states are ballistic~\cite{Mirlin_AA,Mirlin_long,Soumya}.

\section{IV. Mean Square Displacement and Density-density correlation function}
In this section, we calculate the time dependent density-density correlation function $G(x,t)=\langle n_x(t)n_0(t=0)\rangle$.  It gives the probability of finding a particle at site $x$ at time $t$ if initially there was a particle at site $0$. In the infinite temperature limit, $G(x,t)$ is defined as
  \be
  G(x,t) = \frac{1}{Z}\sum_{n}\langle \Phi_n|n_x(t)n_0(t=0)|\Phi_n\rangle
\label{G}
\ee
Here $Z$ is the partition function in the infinite temperature limit and $|\Phi_n\ra$ is an eigenstate of $H$ with eigenvalue $E_n$. We replace the ensemble average in \Eqn{G} by the trace over random states ${|\Psi_r\rangle}$ using the concept of dynamical typicality~\cite{Yevgeny_rev,Reimann} such that
  \be
  G(x,t)=\frac{1}{N_R}\sum_{r=1}^{N_R}\langle \Psi_r|n_x(t)n_0(t=0)|\Psi_r\rangle
  \label{G2}
  \ee
  with $N_R$ being the number of random vectors $\{|\Psi_r\ra\}$ used in the trace. Time evolution is carried out using Chebyshev polynomial method. The resulting correlation function $G(x,t)$ for a given disorder configuration is averaged over many independent disorder configurations to obtain $\overline{G(x,t)}$.  We further calculate the second moment of the correlation function, which is analog to the mean square displacement of a classical particle
  \be
  \langle x^2(t)\rangle = \sum_x x^2 \lbr \overline{G(x,t)} - \overline{G(x,t=0)}\rbr
\label{x2}
  \ee
In the results presented we used number of random initial states in \Eqn{G2} as $N_R=32$ and disorder averaging was done over 50 random configurations for $L=24$. For smaller $L$ values we used $n_R$ upto $80$ and disorder configurations upto $150$.

\Fig{gamma2} shows $\langle x^2(t)\rangle$ as a function of time for $h=5t_0$ and various interaction strengths $V$. After the initial rapid growth, which is common for all the parameters, followed by an oscillatory growth $\mean{x^2(t)}$ saturates for small values of $V$ where the system is fully localized. But for $V\ge 0.3t_0$, $\mean{x^2(t)}$ shows a significant power-law growth $\langle x^2 (t)\rangle \sim t^{\beta}$ in the long time limit. The growth exponent $\beta$ is larger for system with stronger interactions due to enhanced delocalization of states but for all the values of $V$ studied, $\beta \ll 1$ for $h=5t_0$ indicating the existence of a sub-diffusive phase on the delocalized side of the MBL transition point. This is consistent with the analysis of the density imbalance starting from a CDW initial state.
\begin{figure*}[ht]
  \begin{center}
    \vskip0.5cm
    \hspace{-1cm}
  \includegraphics[width=5.8in]
                  {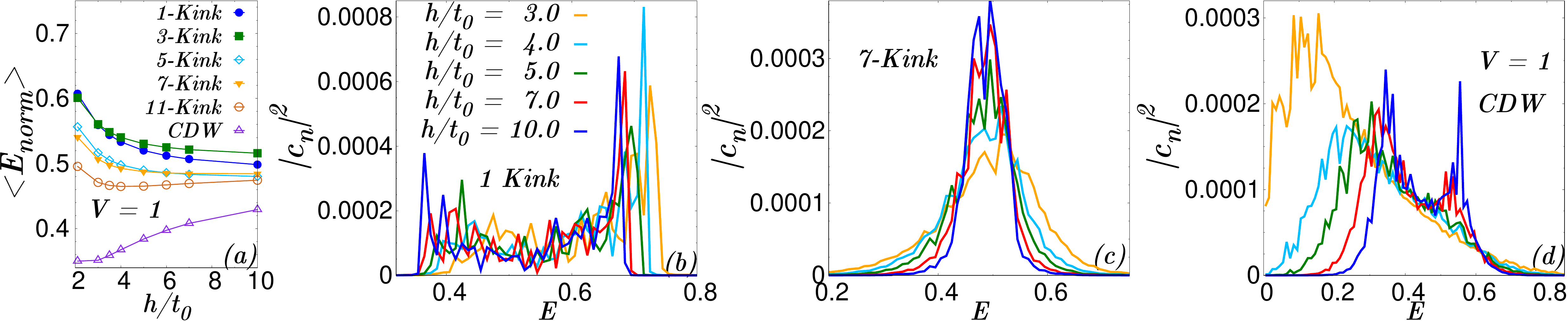}
  \caption{Panel $(a)$ shows normalized energy of the initial state $\la E_{norm} \ra$ vs $h$ for various kink initial states at
           $V = t_0$ and for $L=24$. Panels $(b-d)$ show the overlap $|c_n^2|$ of an initial state with the eigenstates of the Hamiltonian vs the normalized eigen energy $E$  for various values of $h$ for 1-kink, 7-kink and CDW initial states respectively.}
  \label{Ekinks}
\vskip-1cm
\end{center}
\end{figure*}

We compare the exponent $\beta$ obtained from $\mean{x^2(t)}$ with the exponent $\gamma$ obtained from the density imbalance for various kink initial states.  For $V=0$, where the system is fully localized $\beta \sim \gamma\sim 0$ for all the initial states.  As $V$ increases,  the system gets more delocalized and this is indicated in increasing value of $\gamma$ for higher kink states though 1kink and 3-kink states have $\gamma \sim 0$ for all the values of interactions studied here. Thus, for any finite strength of $V$, $\beta/2$ is close to $\gamma$ obtained only from the CDW or other higher kink initial states but $\beta/2 \gg \gamma_{1kink},\gamma_{3kink}$. This is because the long time growth of $\mean{x^2(t)}$ is dominated by extended states while the large time behaviour of the density imbalance is dominated by localized states. Hence, 1-kink state which gets contribution from very few eigenstates shows dynamics much slower than that of $\mean{x^2(t)}$ while for the CDW state the imbalance keeps decaying due to the contribution of a finite fraction of extended states even at long time. Therefore, using 1-kink state to track the MBL to delocalization transition as done in some experimental \cite{Choi2016_Science} as well as in theoretical works~\cite{Mirlin_imb_HF,Imbalance_Knap,Pollmann_DWmelt,Modak2022} will not determine the correct MBL transition point. Lower-kink states will always overestimate the critical $V_c$ for a given disorder strength or will underestimate the critical disorder $h_c$ for a given interaction strength.
\section{V. Energy of various initial states}
At this end, we would like to discuss the energy of various initial states considered in this work. We calculate $E_{in} =\la \Psi_0|H|\Psi_0\ra$ for various kink initial states. In the thermodynamic limit, for any odd-kink state $E_{in} = V\lbr \frac{L}{N_{kinks}+1}-1\rbr$ such that $E_{in}$ is maximum for a 1-kink state and goes to zero for the CDW state.  Even number of kink states are a bit more tricky. An even number of kink state and its particle-hole counterpart state do not have the same energy. For example, 2-kink state as in \Fig{kinks}, has energy equal to the 1-kink state, but the 2-kink state obtained by particle-hole inversion, has energy equal to a 3-kink state. Hence, we have mainly focused on odd number of kink states in this work. \Fig{Ekinks} shows energy of initial states normalized w.r.t the range of eigen-spectrum $E_{norm} = \frac{E_{in}-E_{min}}{E_{max}-E_{min}}$ for each disorder configuration which is then averaged over many independent disorder configurations resulting in $\la E_{norm}\ra$. As shown in \Fig{Ekinks}, 1-kink state has maximum energy due to maximum contribution of the interaction term while the CDW state has the lowest energy. For initial states with 1-kink to 7-kinks, $\la E_{norm}\ra$ shows a slight decrease as the disorder strength increases while for the CDW state $\la E_{norm}\ra$ increases as the disorder strength increases from $2t_0$ to $10t_0$.

Using the expansion of the initial state in terms of eigenstates of the Hamiltonian as done in Section (III), one can write the energy of initial state as a linear combination of eigenstate energies, i.e., $E_{in} = \sum_n |C_n|^2 E_n$. Panels (b-d) of \Fig{Ekinks} show the overlap $|C_n|^2$ of the initial state with the many-body eigenstates of normalised eigen-energy $E$. 1-kink state has larger overlap with eigenstates at the edge of the spectrum. Since the many-body states at the edge of the spectrum gets localized first as the disorder $h$ increases, 1-kink state shows $\gamma \rightarrow 0$ for smaller value of disorder. In contrast, a larger number of kink state, e.g. 7-kink state gets largest contribution from eigenstates in the middle of the spectrum which require largest strength of disorder to get localized. Hence, a 7 kink state shows $\gamma \rightarrow 0$ at larger values of disorder. CDW state is very unique. At small values of disorder $h$, it gets largest contribution from eigenstates at the bottom of the spectrum. As $h$ increases it gets more contribution from states in the middle of the spectrum. In short, eigenstates over a broad energy range contribute to all the initial states studied.
\begin{figure*}[ht]
  \begin{center}
    \vskip0.5cm
    \hspace{-1cm}
  \includegraphics[width=5.1in]
                  {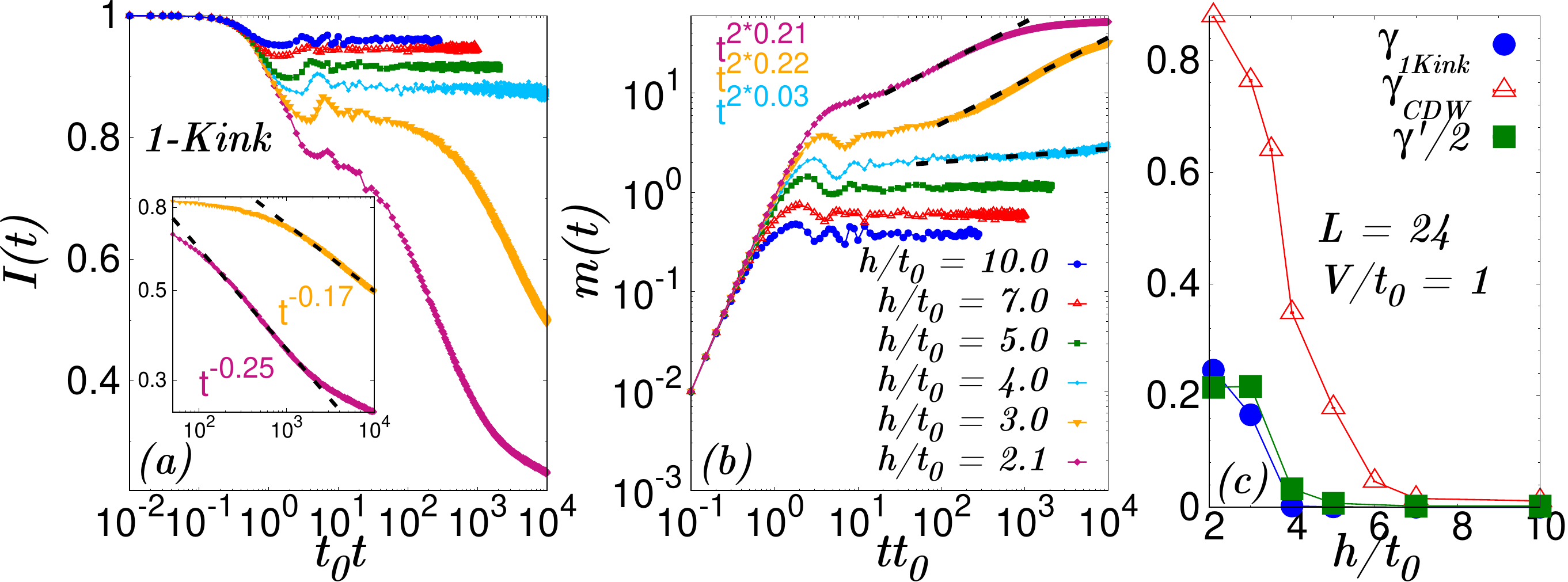}
  \caption{Panel $(a)$ shows the density imbalance $I(t)$ as a function of $t$ for $1-kink$
           initial state while panel $(b)$ shows the first moment of the particle density $m(t)$ across the interface
           of a 1-kink for various strengths of the aperiodic potential $h$
           at $V=t_0$. In panel $(c)$ we show the comparison of exponents
           $\gamma_{{}_{1 Kink}}, \gamma_{{}_{CDW}}$ and $\gamma'/2$.}
  \label{xt}
\vskip-1cm
\end{center}
\end{figure*}

This analysis shows that the initial states considered here do not represent eigenstates in a narrow energy window around the energy of the initial state. Hence, time evolution of various kink initial states can not be used to probe the properties of eigenstates at a specific energy $\la E_{norm}\ra$ as was done in some of the recent works~\cite{Guo_expt_ME,ChandaME}. In fact dynamics of any initial state should not depend on its energy instead it depends on how many eigenstates contribute to it. If an initial state gets contribution from very few eigenstates in a narrow energy window, then the initial state will have slow dynamics while the initial states having contribution from a large number of eigenstates should have faster decay. In this sense the analysis of various initial states presented in this work is very different from the initial states studied recently in context of many-body mobility edges~\cite{Guo_expt_ME,ChandaME}. 

\section{VI. Interface melting for 1-kink state and MBL transition}
In this section, we analyse the melting or broadening of the interface for a 1-kink state as an alternative probe of dynamics. We calculate the following quantities to analyse the broadening of the interface:

\begin{itemize}
  \item The first moment of the particle density $m(t)$ which is defined as 
\be
m(t)=\sum_{i=1}^L i[n_i(t)-n_i(t=0)].
\label{mt}
\ee

By definition, $m(t=0)=0$ and $m(t)$ increases with time. The initial growth is common to all the parameters studied but in the long time limit $m(t)$ has a power-law  growth $t^{\gamma^\prime}$. Since $m(t)$ scales as square of the interface width, $\gamma^\prime$ should be compared with the $2\gamma$. A similar analysis of 1-kink state has been done for a 1-d model with random box disorder case ~\cite{Mirlin_imb_HF} where the dynamical exponent $\gamma^\prime/2$ from 1-kink state was found to be comparable to the exponent $\gamma_{CDW}$ of the density imbalance obtained for a CDW initial state. But, our calculation of $m(t)$ for the aperiodic model leads to a completely different conclusion as shown below.

\item $\Delta N(t)=\sum_{i< L/2} n_i(t)$ which gives the number of particles emitted through the kink or the interface at any time $t$.
\item We also calculate the variance of the particle distribution across the kink
  \be
  Var(t)=\frac{\sum_{i< L/2}i^2 n_i(t)}{\Delta N(t)} -\lbr \frac{\sum_{i<L/2} i n_i(t)}{\Delta N(t)}\rbr ^2
\label{Var}
\ee
Both these quantities, $\Delta N(t)$ and $Var(t)$, have been explored for a 1-d model with random box disorder~\cite{Pollmann_DWmelt}.
\end{itemize}

\begin{figure}[ht]
  \begin{center}
  \includegraphics[width=3.3in]
                  {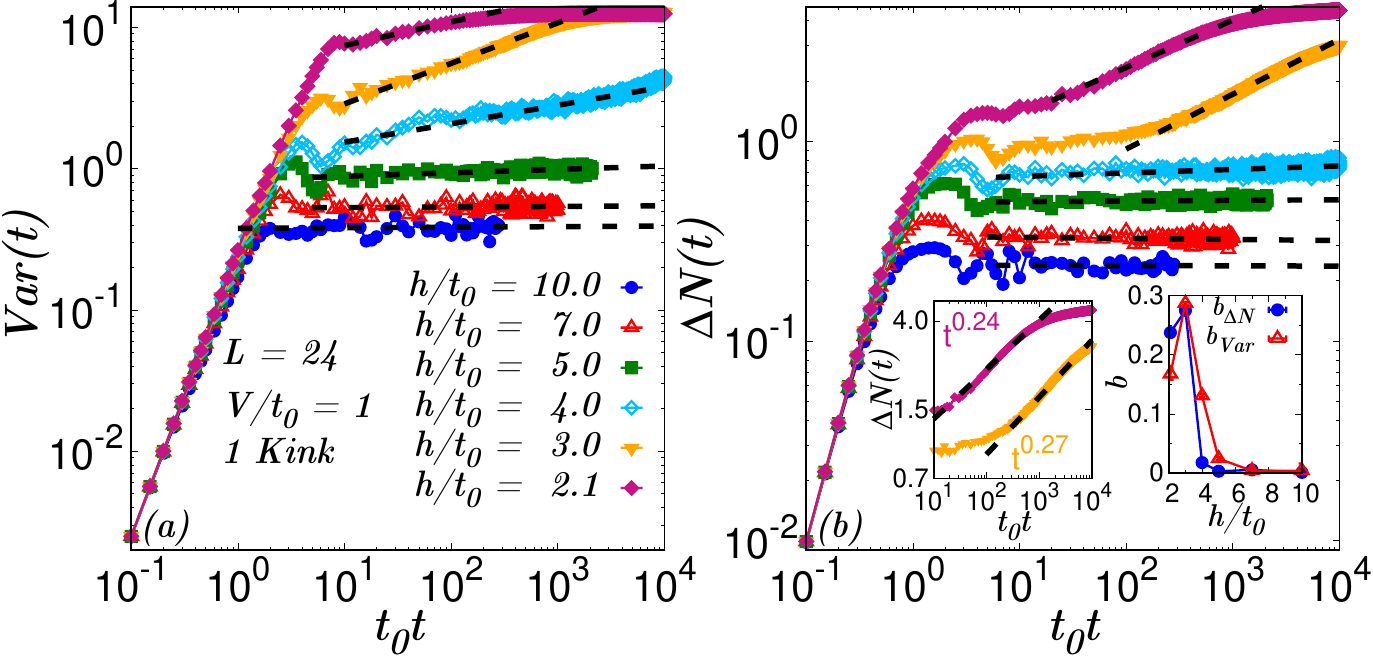}
  \caption{Panel (a) shows the variance of the particle distribution
            and panel $(b)$ presents $\Delta N(t)$ vs $t$ for various values of the aperiodic
            potential $h/t_0$ at $V=t_0$ for $1 kink$ initial state.
           Dashed lines show the fittings of the form $t^b$, 
           and the exponents $b_{\Delta N,Var}$ have been plotted as a function of disorder
           $h/t_0$ in the inset of panel $(b)$.}
  \label{deln}
\vskip-1cm
\end{center}
\end{figure}

In \Fig{xt}, we first compare the dynamics obtained from the time evolution of the density imbalance and the first moment of the particle density $m(t)$ across the interface of a 1-kink state for various strengths of the aperiodic potential $h$ at $V=t_0$.
\begin{figure*}[ht]
  \begin{center}
    \vskip0.5cm
    \hspace{-1cm}
  \includegraphics[width=5.4in]
                  {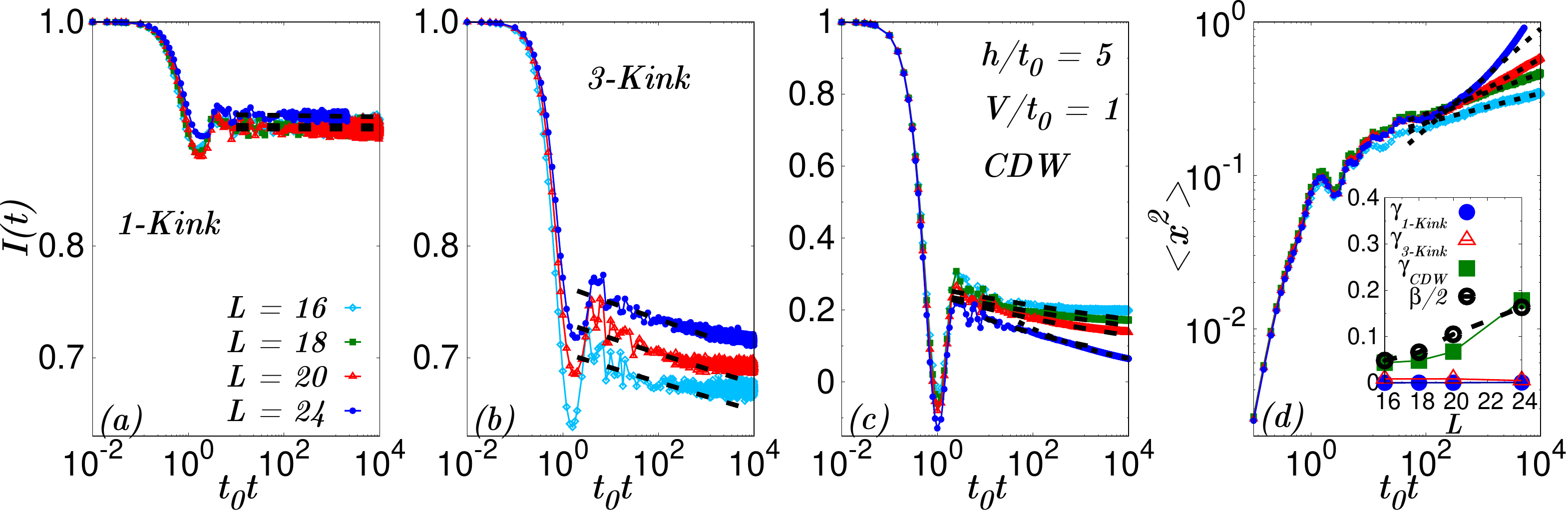} 
  \caption{System size analysis of $I(t)$ for various kink initial states
           at $h=5t_0$, $V=t_0$. For 1-kink and 3-kink initial state,  the
           long-time decay rate of the imbalance decreases as $L$ increases
           resulting in either smaller values of $\gamma$ or larger saturation
           value of imbalance $I_{sat}$. In contrast, for the CDW state the decay gets faster in the long time limit resulting in larger values of $\gamma$ for larger $L$. This is consistent with faster growth of $\mean{x^2(t)}$ in the long time limit.}
  \label{diffL}
\vskip-1.5cm
\end{center}
\end{figure*}

The first moment $m(t)$ shows saturation after initial rapid growth for $h \ge 4t_0$ while $m(t)\sim t^{\gamma^\prime}$ in the long time limit for $h < 4t_0$.  This is completely consistent with $\gamma_{1kink}\rightarrow 0$ at $h=4t_0$. The exponent $\gamma^\prime \sim 2\gamma_{1kink}$ as shown in the right most panel of \Fig{xt}. We would like to stress that on the delocalized side of the MBL transition, both $\gamma_{1kink}$ and $\gamma^\prime/2$ are much less compared to the exponent $\gamma_{CDW}$ obtained from the time evolution of the imbalance starting from a CDW state. Though deep in the MBL phase $\gamma_{1kink}\sim \gamma^\prime\sim \gamma_{CDW} \sim 0$.

Now, we analyse the other two diagnostics of the interface broadening defined above. \Fig{deln} shows $\Delta N(t)$ vs $t$ for various values of the aperiodic potential $h$ at $V=t_0$. At $t=0$, $\Delta N=0$, because all the particles are located in 1st half of the chain. As time increases, more particles get transmitted to the other half of the chain and $\Delta N$ increases with time. For $h \ge 4t_0$, $\Delta N(t)$ saturates after an initial rapid growth, very similar to $m(t)$. For $h < 4t_0$, where almost all the many-body eigenstates are delocalized for $V=t_0$, $\Delta N(t)\sim t^{b_{\Delta N}}$ in the long time limit. The inset in the right panel of \Fig{deln} shows the exponent $b_{\Delta N}$ vs $h$, which has a trend exactly similar to the exponent $\gamma_{1kink}$ obtained from the imbalance $I(t)$ for 1-kink state. We further, analyse the time evolution of the variance $Var(t)$ of the number of particles, shown in the left panel of \Fig{deln}. $Var(t)$ also increases with time, showing saturation after initial rapid growth for larger values of aperiodic potential $h > 4t_0$ and a long time power-law growth $Var(t) \sim t^{b_{Var}}$ appears only for $h < 4t_0$. The exponent $b_{Var} \sim b_{\Delta N} \sim \gamma_{1kink}$ as shown in the inset of the right panel of \Fig{deln}. We would like to stress that this is in contrast to the domain wall analysis done for the random box disorder model~\cite{Pollmann_DWmelt} where the interface melting dynamics has been shown to give a transition point consistent with the level spacing ratio.

The detailed analysis of the interface broadening for the 1-kink initial state further confirms that the dynamics obtained from a 1-kink state, whether it is through the density imbalance or interface broadening, is much slower than the dynamics obtained from the time evolution of the density imbalance starting from a CDW or any other higher kink initial state. In fact on the delocalized side of the MBL transition, the dynamics of a 1-kink and other low kink states is very different from that of the CDW state. This is clearly supported by the system size dependence of the density imbalance for 1-kink and 3-kink states compared with that of a CDW state shown in \Fig{diffL} for $h=5t_0$ and $V=t_0$. For a CDW state, the long-time imbalance shows faster decay resulting in larger values of $\gamma_{CDW}$ as $L$ increases. A similar trend is seen in the system size dependence of $\langle x^2(t)\rangle$ which shows a faster growth in the long time limit for larger values of $L$.
In contrast to this, for 1-kink and 3-kink initial state,  the long-time decay rate of the imbalance decreases as $L$ increases resulting in either smaller values of $\gamma$ or larger saturation value of the imbalance $I_{sat}$.  
\begin{figure*}[ht]
  \begin{center}
    \vskip0.5cm
    \hspace{-1cm}
  \includegraphics[width=6.3in]
                  {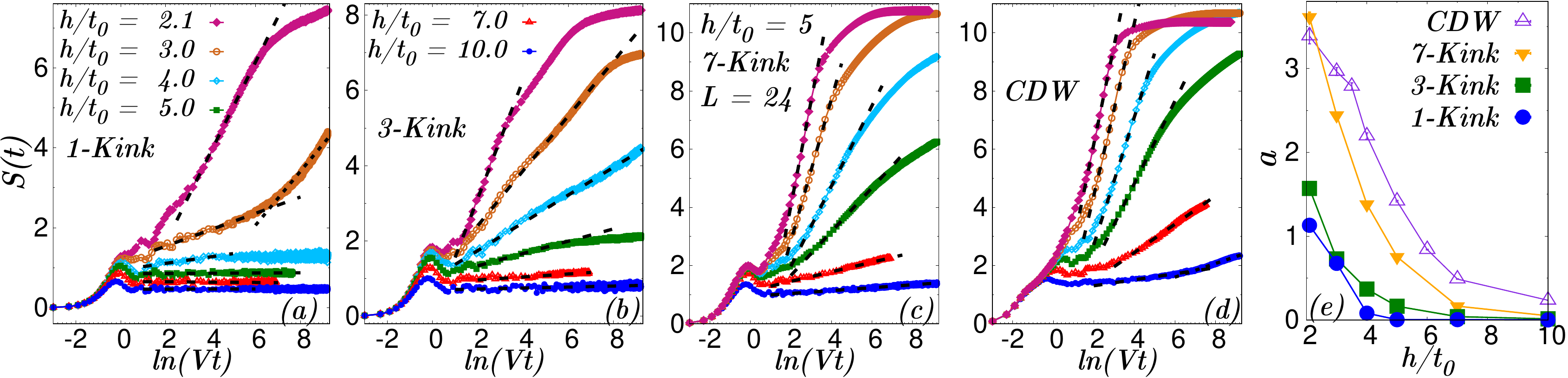}
  \caption{Disorder dependence of the bipartite entanglement entropy
           starting from various kink initial states.
           Panels $(a-d)$ show $S(t)$ as a function of time $t$
           for $1, 3, 7$ kink and $CDW$ initial states for various disorder
           strengths at $V=t_0$ for $L=24$. Dashed lines show the fitting
           to the logarithmic form $S(t) \sim a \ln(Vt)$. Panel $(e)$ shows the coefficient
           $a$ as a function of disorder strength $h/t_0$. Note that $h_c=6.3t_0$ from level spacing ratio (\Fig{rn}).
           }
  \label{SKinks}
\vskip-1cm
\end{center}
\end{figure*}

\section{VII. Kink dependent growth of sublattice entanglement entropy}
We study the growth of sublattice entanglement entropy after a quench from various initial states. We evaluate the bipartite entanglement entropy (EE) by dividing the lattice into two subsystems A and B of sites $L/2$ and study the time evolution of the Renyi entropy $S(t) = -log[Tr_A \rho_A(t)^2]$ where $\rho_A(t)$ is the time evolved reduced density matrix obtained by integrating the total density matrix $\rho_{total}(t) = |\Psi(t)\ra \la \Psi(t)|$ over the degree of freedom of subsystem B. Again, the time evolved state $|\Psi(t)\rangle$ was obtained using Chebyshev method of time evolution.

\Fig{SKinks} shows the bipartite entanglement entropy $S(t)$ as a function of time for various values of $h$ for $V=t_0$ and $L=24$. Let us first understand the growth of EE for a CDW initial state. EE shows a rapid growth at short time which is common to all disorder strengths followed by a logarithmic growth in the long time limit for most of the disorder regime. Not only in the MBL phase, but also in the delocalized phase for weak disorder, $S(t)\sim a \ln(Vt)$ in the long time limit. Though in the entire parameter regime, it is possible to fit $S(t)$ with a $t^c$ form with very small power $c$, the error bars from power-law fit are much larger compared to those for the fit $a \ln(Vt)$.  The coefficient $a$ of the logarithmic growth is larger in the delocalized phase and it decreases as the strength of disorder $h$ increases. Deep inside the MBL phase $a \rightarrow 0$ which is analogous to the Anderson localized phase for which the EE saturates after the initial rapid growth.

We would like to stress that earlier many works have proposed logarithmic growth of EE as a signature of the MBL phase~\cite{Bardarson,Serbyn} but all these works presented exact diagonalization results on very small system sizes ($L\le 12$). Our numerical results show that logarithmic growth of EE seems to be a generic feature of disordered interacting system irrespective of whether the system is delocalized or localized. This puts a question mark on earlier explanations of the logarithmic growth of entanglement entropy in terms of the local integrals of motion which are specific to the MBL phase~\cite{Serbyn}.

\Fig{SKinks} also shows that for any value of the aperiodic potential $h>2t_0$ the coefficient $a$ is smaller for initial states with lower number of kinks and $a$ increases monotonically with $N_{kinks}$. The value of $h$ at which $a \rightarrow 0$ for a 1-kink state is much smaller compared to the corresponding value of $h$ for a larger kink state like CDW state. Interestingly, the strength of aperiodic potential at which $a \rightarrow 0$ is close to $h_{N_{kinks}}$ at which the exponent $\gamma$ from the imbalance goes to zero. In contrast to this, when the system is completely delocalized, e.g. for $h=t_0,V=t_0$ for the model in \Eqn{Ham}, EE shows a logarithmic growth for all the initial states. The short time growth is slower for initial states with lower number of kinks, but  the longer time growth does not necessarily have a systematic dependence on the number of kinks. In short, as long as a part of the eigenspectrum is localized, the growth coefficient $a$ is a monotonic function of the number of kinks in the initial states in complete analogy with the exponent from the density imbalance.

\begin{figure*}[ht]
  \begin{center}
    \vskip0.5cm
    \hspace{-1cm}
  \includegraphics[width=6.5in]
                  {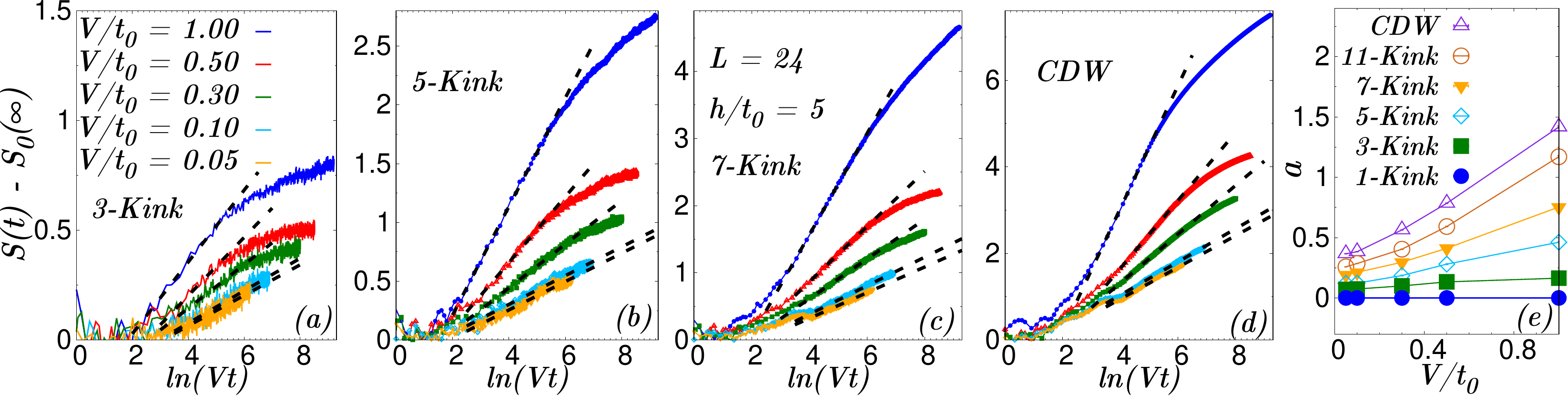}
  \caption{The bipartite entanglement
           entropy $S(t)-S_0(\infty)$ as a function of time $t$ starting from $3, 5, 7$ kink
           and $CDW$ initial states for various interaction strengths at $h=5t_0$
           for $L=24$. Here $S_0(\infty)$ is the saturation value of the EE of the corresponding non-interacting system. Panel $(e)$ shows the coefficient $a$ as a function of
           interaction strength $V/t_0$ starting from various initial states.}
  \label{EEV}
\vskip-1cm
\end{center}
\end{figure*}

We further study the effect of interactions on the EE growth. As the interaction strength, $V$, increases for a fixed value of the aperiodic potential, the system gets more delocalized and hence the rate of growth of EE should increase with $V$. This is what is exactly seen in \Fig{EEV} which shows $S(t)-S_0(\infty)$ vs $Vt$ for $h=5t_0$ and various values of interactions strengths. Here $S_0(\infty)$ is the saturation value of the EE for the corresponding non-interacting system. The most interesting feature to be noticed is that the effect of interactions on the EE growth depends on the initial state started with. A 1-kink state, for $h=5t_0$  shows $a=0$ for all the interactions studied, effectively not showing any logarithmic growth of EE. But for initial states with larger number of kinks, $a(V)$ increases  as $V$ increases such that $\frac{\partial a}{\partial V}$ itself is a monotonically increasing function of the number of kinks in the initial state as shown in panel (d) of \Fig{EEV}.

Hence, 1-kink and other lower kink initial states have a  slower growth of EE than the initial states with larger number of kinks in the long time limit.  This is consistent with our earlier explanation of imbalance decay for low kink states in terms of the suppressed values of $NPR$ of these states. 

\section{VIII. Conclusions and Discussions}
In this work we explored quench dynamics across the MBL transition starting from various initial states characterized by the number of kinks in the density profile. Each of these states gets contribution from a wide range of energy eigenstates of the Hamiltonian under consideration. We showed that the quench dynamics is faster for initial states with large number of kinks, whether it is measured in terms of the time evolution of the density imbalance, or the sublattice entanglement entropy. This is because initial states with larger number of kinks get contributions from a larger fraction of many-body eigenstates of the Hamiltonian. This has interesting effect on the dynamical exponents, rate of growth of various physical quantities and the critical disorder at which the dynamics freezes. Though our study is on a system with aperiodic potential, but we believe that the kink dependence observed in the quench dynamics is very generic and should hold true even for systems with random disorder.

We showed that the dynamical exponent $\gamma$ obtained from the density imbalance increases monotonically with the number of kinks in the initial state, being maximum for the CDW initial state. The exponent $\beta$ from the time evolution of the mean square displacement is consistent with $\gamma$ obtained from the density imbalance only for the CDW and other initial states with large number of kinks such that $\beta \sim 2\gamma$. But $\beta$ is much larger than $\gamma$ obtained from 1-kink and 3-kink initial states. We further showed that the threshold strength of the aperiodic potential, $h_{N_{kinks}}$, at which $\gamma\rightarrow 0$ for a N-kink state is a monotonic function of the number of kinks in the initial state. Interestingly 7-kink state, which has the most symmetric distribution in the eigenbasis of the Hamiltonian around the average energy $E_{norm} \sim 0.5t_0$, shows saturation in the imbalance at $h_{7kink} \sim h_c$ while for the most investigated CDW initial state $h_{CDW} > h_c$. Here $h_c$ is the MBL transition point obtained from the level spacing ratio.

For a 1-kink initial state the critical disorder at which the density imbalance starts showing saturation in the long time limit, $h_{1kink}$, is smallest and is much less than $h_c$. As an alternative probe of dynamics for the 1-kink initial state, we also studied broadening of the kink (or interface) and calculated various measures to quantify this broadening. All these measures provide a dynamics consistent with time evolution of density imbalance for the 1-kink state,  which is much slower than the dynamics of the CDW initial state. Therefore, A 1-kink state always underestimates the critical disorder at which the transition to the MBL phase takes place but it does indicate the onset of a finite fraction of the many-body states or else the imbalance would have decayed to zero. Interestingly, $h_{1kink}$ coincides with the strength of aperiodic potential beyond which the quench dynamics from the CDW initial state as well as the mean square displacement show a subdiffusive phase preceding the MBL phase. This provides an indirect explanation of the subdiffusive phase in this deterministic system in terms of the multifractality of the eigenstates close to the MBL transition~\cite{Luitz2020,Yevgeny-AA,Bera2}.

Our analysis of the entanglement entropy (EE) growth for various initial states showed that the EE shows a logarithmic growth both in the delocalized and the MBL phase. The coefficient of the logarithmic term is a monotonically increasing function of the number of kinks in the initial state and goes to zero for $h > h_{N_{kinks}}$ showing a complete consistency with the time evolution of the density imbalance. Further, the coefficient of logarithmic growth increases with increase in the interactions and the rate of growth with interactions itself is a function of number of kinks in the initial states.It will be useful to check the observations made in this work by studying larger system sizes using methods like time dependent variational principle. It will also be interesting to have an experimental verification of our numerical results on various kink initial states in an optical lattice experiment.

\section{Acknowledgements}
A.G. would like to acknowledge Science and Engineering Research Board (SERB) of Department of Science and Technology (DST), India under grant No. CRG/2018/003269 for financial support. Y.P. would like to acknowledge DST for funding, and SINP cluster facilities.
A.G. also acknowledges National Supercomputing Mission (NSM) for providing computing resources of ‘PARAM Shakti’ at IIT Kharagpur, which is implemented by C-DAC and supported by the Ministry of Electronics and Information Technology (MeitY) and Department of Science and Technology (DST), Government of India.
\section{Appendix A. Statistics of Level spacing ratio}
\begin{figure}[ht]
  \begin{center}
  \includegraphics[width=2.5in]
                  {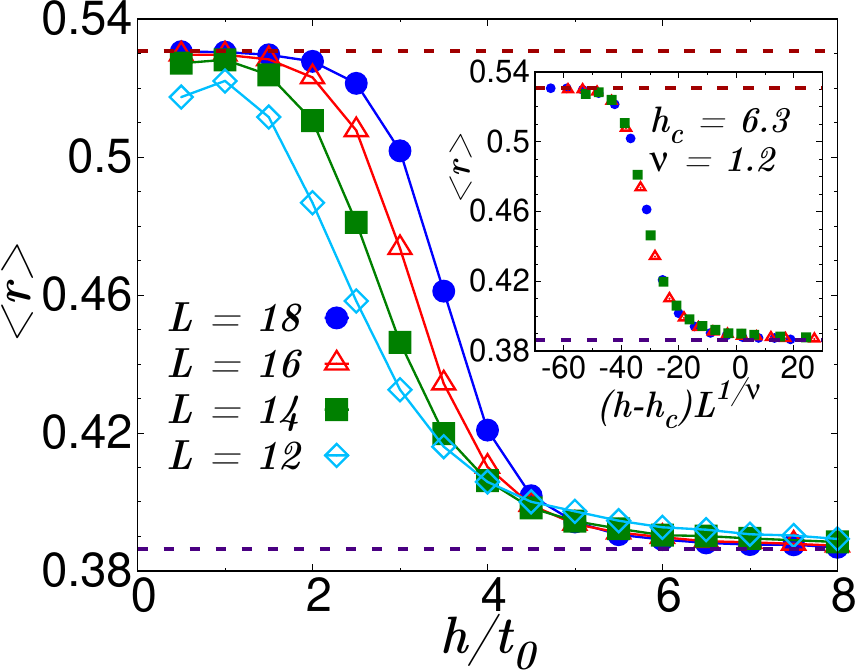}
  \caption{ Energy level spacing ratio of successive gaps $\mean{r}$ vs disorder $h$.
            $\mean{r}$ is obtained by averaging $r_n$ over the entire energy spectrum and over a large number of independent disorder configurations. Inset shows the
            data collapse to obatin the critical disorder $h_c \sim 6.3t_0$ with exponent $\nu \sim 1.2$. The data have been averaged over $10000-150$ configurations for $L = 12-18$.}
  \label{rn}
\vskip-1cm
\end{center}
\end{figure}
To obtain the critical disorder $h_c$ at which delocalization to MBL transition takes place for $V=t_0$, we solved the Hamiltonian in \Eqn{Ham} using exact diagonalization for system sizes upto 18 sites and obtained the eigenvalues $E_n$ for every disorder configuration. We calculated the level spacing ratio $r_n=\delta_n/\delta_{n+1}$ with $\delta_n=E_{n+1}-E_n$ and average it over the entire spectrum as well as over a large number of independent disorder configurations for several system sizes to obtain average $\mean{r}$ shown in \Fig{rn}. For small values of $h$, $\mean{r}$ increases with the system size approaching the Wigner-Dyson value of $0.529$; while for strong disorder $\mean{r}$ decreases for larger values of $L$  approaching the Poissonian value of $0.389$~\cite{Alet_rev}.  The data collapse showed $h_c \sim 6.3t_0 $ and the exponent $\nu \sim 1.2$.
\section{Appendix B. Time Evolution of Imbalance for non-interacting case}
\begin{figure}[ht]
  \begin{center}
  \includegraphics[width=3.4in]
                  {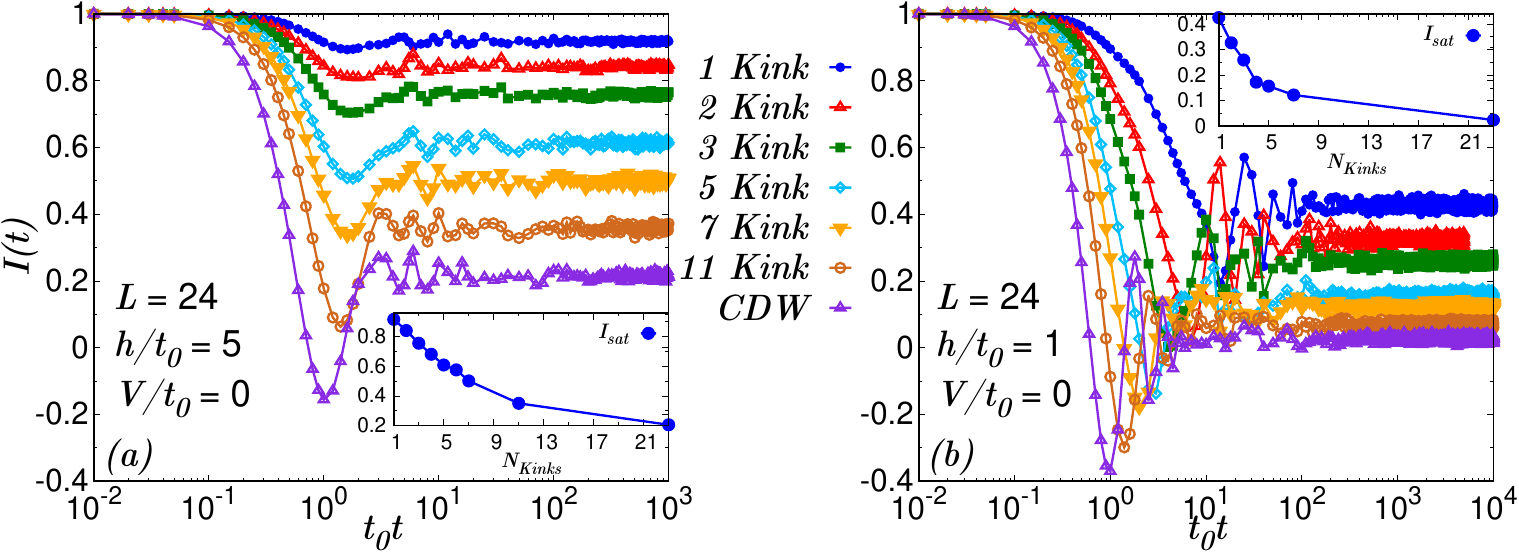}
  \caption{Density imbalance $I(t)$ in the localized and
           delocalized regime of the non-interacting system.
           Panel $(a)$ shows $I(t)$ as a function of $t$ starting
           from various kink initial states at $h=5t_0$ for the non-interacting system for $L=24$. Dynamics
           depends strongly on the initial states. The saturation
           value $I_{sat}$, shown in insets, decreases as the number
           of kinks increases in the initial states. Panel (b) shows the density imbalance for $h=t_0$ where the non-interacting system has single particle mobility edges.}
  \label{AL}
\vskip-1cm
\end{center}
\end{figure}
As shown in \Fig{AL}, for $V=0$ and $h=5t_0$, where all the single particle states are localized, $I(t)$ shows saturation in the long time limit similar to the imbalance in the MBL phase. $I_{sat}$ increases with $N_{kinks}$ in the initial state, being largest for the 1-kink state and smallest for the CDW state. Even more interestingly, for $h=t_0$, where the non-interacting system has single particle mobility edges at $\pm |2t_0-h|$, the imbalance saturates in the long time limit. The difference compared to the fully localized case is that now $I_{sat} \sim 0$ for larger kink states including the CDW state. Thus, even if one can not distinguish between an Anderson localized phase and MBL phase through time evolution of imbalance, the approach to the localized phase as disorder $h$ is increased can distinguish between the two localized phases. The delocalised side of the MBL phase has a power-law decay of imbalance while the delocalized side of Anderson localized phase does not show any power-law decay of imbalance. 
\section{Appendix C. Density Imbalance for a completely delocalized phase}
  
\begin{figure}[ht]
  \begin{center}
  \includegraphics[width=3.5in]
                  {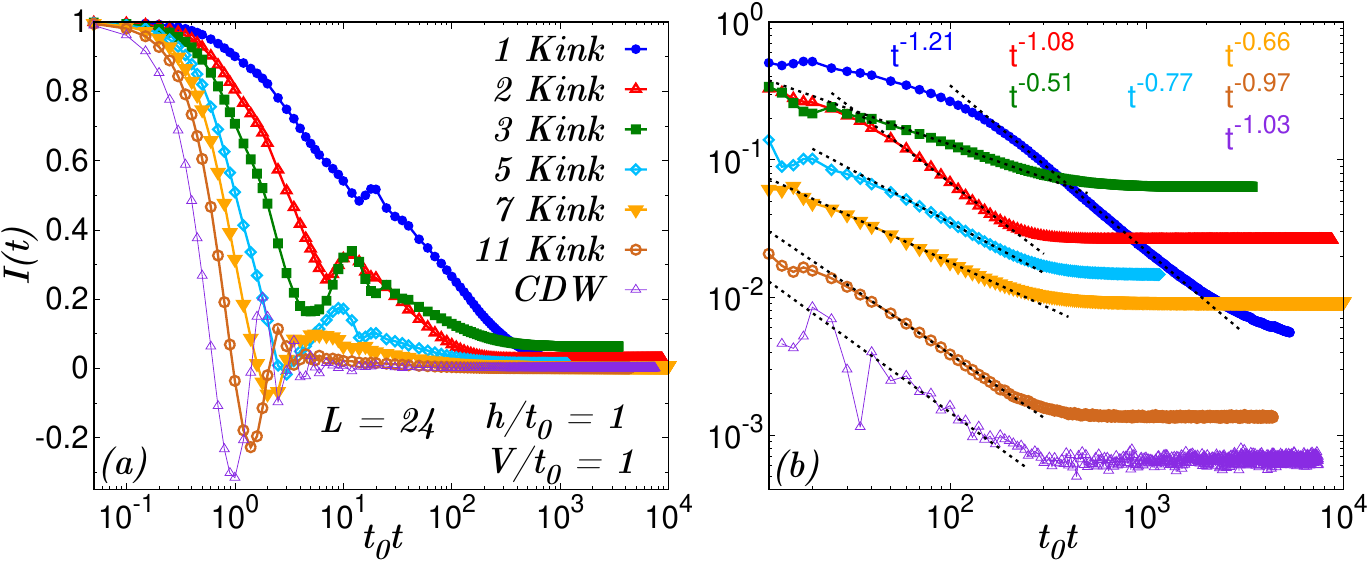}        
  \caption{Density imbalance $I(t)$ in the fully delocalized regime.
           Panel $(a)$ shows $I(t)$ as a function of $t$ starting
           from various kink initial states at $h=t_0, V=t_0$ for
           $L=24$. The long time fitting has been shown in
           panel $(b)$ where power-law fits have been shown by dashed
           lines. In the delocalized regime there is
           no monotonic increase in $\gamma$ with the number of kinks
           in the initial states.}
  \label{h1}
\vskip-1cm
\end{center}
\end{figure}
\begin{figure}[ht]
  \begin{center}
    \vskip0.5cm
    \hspace{-1cm}
  \includegraphics[width=2.1in]
                  {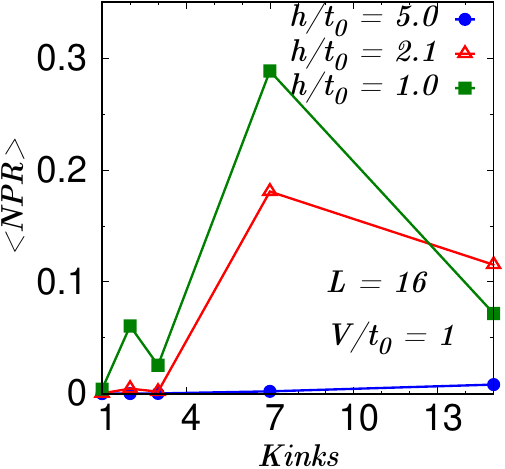}             
  \caption{NPR of various initial states as a function of number of kinks in the initial state at $V=t_0$ for $L=16$ sites chain.
           For $h=t_0$ and $h=2.1t_0$, $NPR$ does not have a systematic dependence on the number of kinks $N_{kinks}$ in initial states. This should be compared with a systematic dependence of $NPR$ with $N_{kinks}$ for $h=5t_0$. }
  \label{nprh1}
\vskip-1cm
\end{center}
\end{figure}

In this Appendix we discuss the time evolution of the density imbalance starting from various kink initial states for $h=t_0$ and $V=t_0$ where the system is completely delocalized. In the absence of interactions, the system has single particle mobility edges at $h=t_0$ at $\pm|h-2t_0|$ and all the many-body eigenstates of the half-filled system are delocalized and ballistic. \Fig{h1} shows that the initial short time decay of $I(t)$ is slower for initial states with less number of kinks. This can be explained in terms of the lower connectivity of initial states with less number of kinks in the Fock space of a model with nearest neighbour hopping. But in the long time limit, the dynamics become fast due to completely delocalized states and the imbalance $I(t) \rightarrow 0$ for all the kink states. Note that for this case of fully delocalized states $\gamma$ does not have a systematic dependence on the number of kinks in the initial states. This is also consistent with the fact that $NPR$ of initial states with various number of kinks is not a monotonic function of $N_{kinks}$ for $h < 2t_0$ as shown in \Fig{nprh1}. In contrast to this for $h>2t_0$, $\gamma$ has a monotonic dependence on $N_{kinks}$ which is directly related with increase of $NPR$ of initial state with $N_{kinks}$ as shown in the main text.  

\bibliography{bibliography_12June}

\end{document}